%% file: main.tex
\documentclass[sigconf, nonacm]{acmart}

\usepackage{textcase}
\makeatletter
\expandafter\def\expandafter\@secfont\expandafter{\@secfont\MakeTextUppercase}
\makeatother

\usepackage{algorithmic}

\usepackage{array}
\usepackage[skins]{tcolorbox}
\usepackage{caption}
\usepackage[ruled,vlined]{algorithm2e}
\usepackage{booktabs}
\usepackage{pifont}
\usepackage{color}
\usepackage{soul, framed}
\usepackage{makecell}
\usepackage{xcolor}
\usepackage{colortbl}
\usepackage{graphicx}
\usepackage{textcomp}
\usepackage{xcolor,colortbl}
\usepackage{multirow}
\usepackage{bm,dsfont}
\usepackage{subfigure}
\usepackage{balance}
\usepackage{stmaryrd}
\usepackage{autobreak}
\usepackage{enumitem}
\usepackage{bbding}
\usepackage{pifont}
\usepackage{wasysym}

\usepackage{amssymb}
\usepackage{booktabs}
\usepackage{multirow}
\usepackage{graphicx}
\usepackage{xcolor}
\usepackage{colortbl}

\definecolor{sotabg}{HTML}{FCE8E8}
\definecolor{gainbg}{HTML}{E7F0F8}

\newcommand{\sotacell}[1]{%
  \cellcolor{sotabg}\ensuremath{\mathbf{#1}}%
}

\newcommand{\gaincell}[1]{%
  \cellcolor{gainbg}\ensuremath{\mathbf{#1}}%
}

\newcommand{\our}{\emph{REMIX}}

\newtheorem{definition}{Definition}

\newtheorem{example}{Example}

\usepackage[skins]{tcolorbox}
\usepackage{enumitem}

\newtcolorbox[auto counter]{promptbox}[1][]{
    enhanced,
    width=\linewidth,
    title={Prompt Design},
    colback=black!2,
    colframe=black!60,
    colbacktitle=black!60,
    coltitle=white,
    fonttitle=\bfseries\normalsize,
    fontupper=\small,
    boxrule=0.8pt,
    titlerule=0pt,
    arc=2mm,
    outer arc=2mm,
    boxsep=0pt,
    left=8pt,
    right=8pt,
    top=6pt,
    bottom=6pt,
    lefttitle=8pt,
    righttitle=8pt,
    toptitle=4pt,
    bottomtitle=4pt,
    before skip=6pt,
    after skip=6pt,
    #1
}

\begin{document}
% \title{Synthesizing Multi-Table Relational Databases with Free-Text Attributes via Graph-Conditioned Language Models}
% \title{GRACE: Mixed-Type Relational Database Synthesis with
% % Graph-Guided Language Models and Constraint Enforcement}
% \title{Graph-Prompted Language Models Are Realistic Generators of Mixed-Type Relational Databases}
\title{Mixed-Type Relational Database Synthesis with Graph-Prompted Language Model}

\author{Yuxiang Sun}
% \authornote{Yuxiang Sun and Mengshi Chen contributed equally to this
% work and share first authorship.}
\affiliation{%
  \institution{Shanghai Jiao Tong University}
  \city{Shanghai}
  \country{China}
}
\email{scscsc04@sjtu.edu.cn}

\author{Mengshi Chen}
% \authornotemark[1]
\affiliation{%
  \institution{Shanghai Jiao Tong University}
  \city{Shanghai}
  \country{China}
}
\email{chenmengshi@sjtu.edu.cn}

\author{Jianwei Wang}
\affiliation{%
  \institution{University of New South Wales}
  \city{Sydney}
  \country{Australia}
}
\email{jianwei.wang1@unsw.edu.au}

\author{Wenjie Zhang}
\affiliation{%
  \institution{University of New South Wales}
  \city{Sydney}
  \country{Australia}
}
\email{wenjie.zhang@unsw.edu.au}

\author{Kai Wang}
\affiliation{%
  \institution{Shanghai Jiao Tong University}
  \city{Shanghai}
  \country{China}
}
\email{w.kai@sjtu.edu.cn}

\begin{abstract}
Data access is often limited by privacy or proprietary restrictions, motivating relational database synthesis, which synthesizes realistic contents across related tables.
Relational databases couple multi-table structure with mixed-type attributes and intricate semantic dependencies among values, making realistic synthesis challenging.
% However, most existing methods use statistical or neural models tailored to numerical and categorical attributes, leaving textual attributes largely unsupported, while probabilistic generation can violate semantic dependencies.
In this paper, we propose \our{}, a unified framework for realistic \textbf{\underline{RE}}lational database synthesis with \textbf{\underline{MIX}}ed-type attributes. \our{} integrates a small language model (SLM) and a large
language model (LLM) through two key phases:
\emph{Graph-to-Prompt Tuning} and
\emph{Collaborative SLM--LLM Decoding}.
To capture attribute dependencies within and across tables, \emph{Graph-to-Prompt Tuning} represents the database as an attributed heterogeneous graph.
We project mixed-type record embeddings and aggregated relational context into soft prompts that serve as continuous vectors for an SLM.
We jointly optimize the graph-to-prompt mapping and SLM through autoregressive attribute reconstruction.
Furthermore, to reduce semantic inconsistencies, \emph{Collaborative SLM--LLM Decoding} combines SLM-based record generation with repair programs induced by an LLM.
The LLM discovers semantic dependencies, iteratively
refines them through independent validation with a statistical
bound on the overall probability of false acceptance, and translates validated dependencies into tested programs that repair generated values.
Extensive experiments on seven real-world relational databases demonstrate the effectiveness of \our{} in generating high-fidelity data with improved semantic consistency and downstream utility.
\end{abstract}

\maketitle

\pagestyle{plain}

\input{introduction}
\input{preliminary}
\input{tuning}
\input{decoding}
\input{experiment}
\input{relatedwork}
\input{conclusion}

\newpage
\bibliographystyle{ACM-Reference-Format}
\bibliography{sample}

\iffalse
\newpage
\appendix
\onecolumn
\section{Appendix}
\input{appendix.tex}
\label{sec.appendix}
 \fi 

\end{document}

%% file: introduction.tex
\section{INTRODUCTION}
\label{sec:introduction}

\begin{figure}[!t]
    \centering
    \includegraphics[width=\columnwidth]{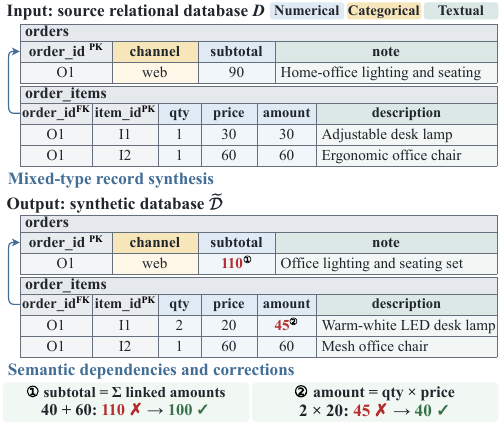}
    \captionsetup{skip=3pt}
    \caption{Mixed-type relational database synthesis.}
    \label{fig:motivation-example}
\end{figure}

\begingroup
\setlength{\textfloatsep}{15pt}
\setlength{\dbltextfloatsep}{12pt}
\begin{figure*}[t]
    \centering
    \includegraphics[width=\textwidth]{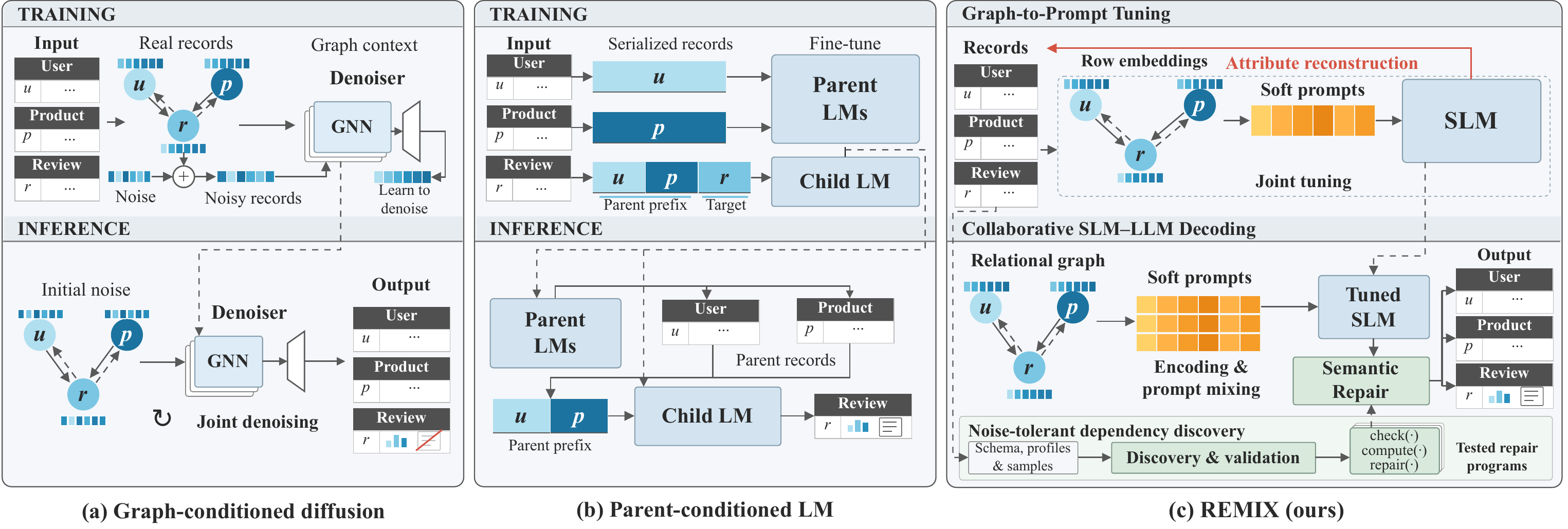}
    \caption{Framework comparison for relational database synthesis.
    (a) Graph-conditioned diffusion trains a GNN denoiser with relational context, then jointly denoises related records from noise.
    (b) Parent-conditioned LM trains on serialized records, then generates parents followed by parent-conditioned children.
    (c) \our{} jointly tunes its graph-to-prompt mapping and SLM via attribute reconstruction, while an LLM prepares tested programs from validated semantic dependencies. During decoding, the SLM generates mixed-type records from soft prompts, and the programs compute derived values and repair inconsistencies.}
    \label{fig:framework-comparison}
\end{figure*}

Relational databases organize records and represent their relationships across tables~\cite{codd}. They support efficient data management and retrieval, making them a cornerstone of large-scale information systems in e-commerce~\cite{ecommerce, ecommerce2}, healthcare~\cite{health}, and enterprise applications~\cite{enterprise,enterprise2}. However, access to real-world databases is often limited by privacy concerns, proprietary restrictions, and data collection costs~\cite{privacy1, privacy2,privacy3}. Relational database synthesis addresses this problem by generating synthetic database content that preserves the statistical properties and relational dependencies of the source data~\cite{sdv, syntherela}, thereby providing an alternative data source for downstream applications.

Real-world relational databases contain numerical, categorical, and textual attributes whose dependencies can span multiple tables~\cite{text1,relbench,text3}.
Textual descriptions provide contextual details that complement numerical and categorical values.
Meanwhile, semantic dependencies express arithmetic relationships, domain constraints, and business rules at both record and database levels.
Excluding attributes or failing to preserve their semantic dependencies can therefore affect analytical results and downstream predictions.
Figure~\ref{fig:motivation-example} illustrates this synthesis task using a shopping database, where an item's amount equals its quantity times its unit price and an order's subtotal equals the sum of its linked item amounts.

\begin{table}[!t]
\caption{%
  Comparison of relational database synthesis methods.
  \ding{51} denotes support; \ding{55} denotes no native support.
  Abbr: AR (autoregressive), GAN (generative adversarial network),
  GNN (graph neural network), LM (language model).
}
\label{tab:comparison}
\centering
\scriptsize
\setlength{\tabcolsep}{3.0pt}
\renewcommand{\arraystretch}{1.00}
\resizebox{\columnwidth}{!}{%
\begin{tabular}{l ccc c c}
\toprule
\multirow[c]{2}{*}[-0.5ex]{\textbf{Method}}
  & \multicolumn{3}{c}{\textbf{Column Type Coverage}}
  & \multirow[c]{2}{*}[-0.5ex]{%
      \textbf{\shortstack{Semantic\\Repair}}}
  & \multirow[c]{2}{*}[-0.5ex]{\textbf{Backbone}}
\\
\cmidrule(lr){2-4}
  & \textbf{Num.} & \textbf{Cat.} & \textbf{Text} & & \\
\midrule
SDV~\cite{sdv}
  & \ding{51} & \ding{51} & \ding{55}
  & \ding{55} & Statistical Model \\
RC-TGAN~\cite{rctgan}
  & \ding{51} & \ding{51} & \ding{55}
  & \ding{55} & GAN \\
TabularARGN~\cite{argn}
  & \ding{51} & \ding{51} & \ding{55}
  & \ding{55} & AR \\
GRDM~\cite{grdm}
  & \ding{51} & \ding{51} & \ding{55}
  & \ding{55} & GNN+Diffusion \\
REaLTabFormer~\cite{realtabformer}
  & \ding{51} & \ding{51} & \ding{55}
  & \ding{55} & LM \\
\midrule
\textbf{\our{}}
  & \ding{51} & \ding{51} & \ding{51}
  & \ding{51} & \textbf{GNN+LM} \\
\bottomrule
\end{tabular}%
}
\end{table}

Table~\ref{tab:comparison} compares representative relational database synthesis methods. In early stages, SDV~\cite{sdv} synthesizes related tables using a hierarchical statistical framework, but its assumptions and aggregate summaries can limit the dependencies captured. Conventional neural methods model attribute distributions and relational dependencies through generative adversarial networks (GAN) in RC-TGAN~\cite{rctgan}, variational autoencoders (VAE) in GraphVAE~\cite{graphvae}, and autoregressive (AR) models in TabularARGN~\cite{argn}. However, GANs can suffer from instability and mode collapse, VAEs balance regularization and reconstruction fidelity, and AR models can accumulate errors. Their native support is generally limited to numerical and categorical attributes.

% More recently, the success of diffusion models in high-fidelity generation and language models (LM) in capturing contextual dependencies has motivated their adoption for relational database synthesis. 
Recent methods apply diffusion models and language models (LM) to relational database synthesis.
As depicted in Figure~\ref{fig:framework-comparison}(a), diffusion-based methods such as GRDM~\cite{grdm} and RelDiff~\cite{reldiff} jointly denoise related records conditioned on relational context to capture cross-table dependencies. In parallel, LM-based synthesizers serialize records into token sequences and generate new records autoregressively. For example, REaLTabFormer~\cite{realtabformer} generates child records conditioned on previously generated parent records, as depicted in Figure~\ref{fig:framework-comparison}(b). GReaTER~\cite{greater} instead restructures and joins selected related records before serialization and generation. 

\noindent\textbf{Motivations.}
LM-based synthesis can generate structured values and text through a common token representation~\cite{great}.
Existing relational LM methods, however, typically obtain relational context from serialized parent records or selected table joins~\cite{realtabformer,greater}, limiting the related information available for joint generation.
Relational diffusion models incorporate cross-table context but primarily target numerical and categorical attributes~\cite{grdm}.
Generating text independently of these attributes can overlook their dependencies.
These limitations motivate joint modeling of mixed-type attributes across related tables.

{\emergencystretch=1em
Moreover, a synthetic database can closely match individual attribute distributions while violating semantic dependencies~\cite{cvr1}.
For example, in a shopping database, an order's \texttt{subtotal} should equal the sum of its linked item amounts.
A synthesized order with item amounts of 40 and 60 but a subtotal of 110 violates this dependency despite valid foreign-key links.
A naive approach is to check generated records directly, either manually or with an LLM, but such checks can overlook dependencies and yield inconsistent judgments across repeated evaluations.
These limitations motivate incorporating semantic dependencies into synthesis to reduce inconsistencies among generated values.
\par}

\noindent\textbf{Challenges.}
Modeling mixed-type attributes across related tables while reducing semantic inconsistencies involves two challenges.

\noindent\textit{Challenge I.}
Incorporating relational context into joint modeling of mixed-type attributes is difficult because relevant information spans heterogeneous records and multiple relationship paths.
Parent conditioning can omit information from other connected records, while one-to-many joins repeat shared records and produce long serialized inputs.
Compressing input while preserving the attribute semantics and relation types needed to capture cross-table dependencies is therefore difficult.

\noindent\textit{Challenge II.}
{
The diverse forms and context-dependent applicability conditions of semantic dependencies complicate their reliable discovery from source data and safe application during synthesis.
% Semantic dependencies include arithmetic equalities, functional dependencies, and domain-specific rules.
Inferred dependencies may reflect statistical associations or hold only under specific conditions.
Applying them to all synthesized records can therefore lead to incorrect changes.
Moreover, noise in the source databases can also cause occasional violations of otherwise valid dependencies, making it difficult to distinguish unreliable rules from valid ones affected by noise.
%When dependencies share attributes, changing a value to satisfy one can violate another.
%Enforcing semantic dependencies without introducing new violations or distorting statistical dependencies is difficult.
}

\noindent\textbf{Our approaches.}
To address these challenges, we propose \our{}, a unified framework for mixed-type relational database synthesis, as illustrated in Figure~\ref{fig:framework-comparison}(c).
It first trains a small language model (SLM) for graph-conditioned generation through \emph{Graph-to-Prompt Tuning}, then combines this generator with repair programs induced by a large language model (LLM) in \emph{Collaborative SLM--LLM Decoding}.

To address Challenge I, \emph{Graph-to-Prompt Tuning} learns to supply mixed-type record information and relational context through soft prompts, continuous input vectors that condition the SLM.
Specifically, we represent records and foreign-key links as an attributed heterogeneous graph.
A type-aware row encoder maps each record's mixed-type attributes to a row embedding, while a heterogeneous graph neural network (GNN) aggregates information from connected records.
A projector maps the relational context together with a randomly masked row embedding into soft prompts.
Masking encourages the SLM to use relational context during reconstruction.
An autoregressive attribute reconstruction objective jointly trains the graph-to-prompt mapping and the SLM.
This design supplies relational context without serializing joined tables or requiring a fixed table generation order.

To address Challenge II, \emph{Collaborative SLM--LLM Decoding} combines the tuned SLM generator with programs constructed by the LLM to  repair violations of semantic dependencies.
%The LLM first proposes semantic dependencies from source data.
%Independent validation provides counterexamples for refining these dependencies and their applicability conditions, while a statistical acceptance criterion bounds the overall probability of false acceptance.
The LLM first proposes semantic dependencies from source data. Validation on independent samples accounts for source noise through a small violation tolerance and uses counterexamples to guide iterative refinement. A statistical acceptance criterion bounds the overall probability of accepting dependencies whose violation probabilities exceed the prescribed tolerance.
The LLM then translates validated dependencies into executable programs, which are tested individually and together.
The tested programs are reused across generated records.
After the tuned SLM generates new records, the programs execute in dependency order to  repair violations.
\our{} checks interacting program updates together and rolls back changes that introduce new violations.

\noindent\textbf{Contributions.}
Our main contributions are summarized as follows.
\begin{itemize}[leftmargin=*, labelindent=0pt]
\item We study mixed-type relational database synthesis and propose \our{}, a unified framework for jointly generating structured values and text across related tables.
\item We design \emph{Graph-to-Prompt Tuning}, which uses type-aware encoding and heterogeneous graph propagation to construct soft prompts for learning mixed-type attribute distributions.
\item We develop \emph{Collaborative SLM--LLM Decoding}, combining SLM generation with independently validated semantic dependencies and tested programs built by an LLM for computation and repair.
% The following figures are retained from introduction (1).tex.
% Check the repaired CVR before submission: the repository's current
% seven-generation-seed report gives 9.04%, whereas that draft gives 7.67%.
\item Experiments on seven real-world databases show \our{}'s gains in fidelity, semantic consistency, and downstream utility.
\end{itemize}
\endgroup

%% file: preliminary.tex
\section{PRELIMINARIES}
\label{sec:preliminaries}

\begin{table}[t]
    \captionsetup{skip=3pt}
    \caption{Symbols and Descriptions}
    \label{tab:notation}
    \centering
    \small
    \setlength{\tabcolsep}{4pt}
    \renewcommand{\arraystretch}{1.10}
    \begin{tabular}{|
        >{\raggedright\arraybackslash}p{0.22\columnwidth}|
        >{\raggedright\arraybackslash}p{\dimexpr
            0.78\columnwidth-4\tabcolsep-3\arrayrulewidth\relax}|}
        \hline
        \rowcolor[gray]{0.8}
        \textbf{Notation} & \textbf{Description} \\
        \hline
        \rule{0pt}{2.8ex}$\mathcal{D},\widetilde{\mathcal{D}}$
        & Source and synthesized relational databases \\
        \hline
        $\mathcal{R},\mathcal{L}$
        & Tables and foreign-key relationships \\
        \hline
        $\mathcal{G},\mathcal{V},\mathcal{E}$
        & Attributed graph, record nodes, and typed edges \\
        \hline
        $\mathbf{r}_v,\widetilde{\mathbf{r}}_v$
        & Row embedding and its randomly masked version \\
        \hline
        $\mathbf{h}_v$
        & GNN-derived context embedding \\
        \hline
        $\mathbf{P}_v$
        & Soft prompt used to condition the SLM decoder \\
        \hline
        $q_{\phi(v)}$
        & Table-specific generation instruction \\
        \hline
        $\mathbf{y}_v,\widetilde{\mathbf{y}}_v$
        & Serialized training label and generated sequence \\
        \hline
        $\Theta$
        & Trainable generator parameters \\
        \hline
        $\mathcal{C},\mathcal{C}^{*}$
        & Human-verified reference / LLM-discovered semantic dependency sets \\
        \hline
        $\mathcal{F}$
        & Semantic repair functions \\
        \hline
    \end{tabular}

\end{table}

\subsection{Problem Statement}
\label{sec:problem_statement}

We represent a relational database as $\mathcal{D}=(\mathcal{R},\mathcal{L})$, where $\mathcal{R}=\{R^{(1)},\ldots,R^{(m)}\}$ is a collection of $m$ tables and $\mathcal{L}$ is the set of primary-key--foreign-key relationships.
Each table $R^{(a)}$ contains $n_a$ records with numerical, categorical, or textual non-key attributes.
Primary keys uniquely identify records within a table, while foreign keys link records across tables.

\begin{definition}[Relational Database Synthesis]
\label{def:relational_database_synthesis}
Given a source relational database $\mathcal{D}$, relational database synthesis aims to generate a synthetic database $\widetilde{\mathcal{D}}$ by synthesizing numerical, categorical, and textual non-key attribute values while preserving the source schema, table sizes, and primary- and foreign-key assignments.
The goal is to reproduce attribute distributions and statistical relationships within and across tables as faithfully as possible.
\end{definition}

In this paper, we focus on synthesizing database contents while preserving the source schema, table sizes, and key assignments.
Beyond matching statistical distributions, realistic synthesis requires preserving semantic dependencies among attribute values, which are introduced in Section~\ref{sec:semantic_dependencies}.
To jointly model and generate numerical, categorical, and textual attributes, language models provide a common autoregressive formulation.
Section~\ref{sec:prelim_llm} presents this formulation and the adaptation techniques used in our approach.

\subsection{Semantic Dependencies}
\label{sec:semantic_dependencies}

Database content reflects real-world knowledge and application semantics, which impose requirements on how attribute values relate within and across tables.
These requirements often remain implicit in the data and extend beyond declared schema constraints.
In this work, semantic dependencies encompass functional dependencies, arithmetic relationships, and domain-specific rules.
These dependencies are defined in Definition~\ref{def:semantic_dependency} and illustrated in Example~\ref{ex:semantic_dependencies}.

\begin{definition}[Semantic Dependency]
\label{def:semantic_dependency}
A semantic dependency is a deterministic rule over attribute values, specified by (i) a scope identifying the records or record groups to be checked within or across tables, (ii) an applicability condition, and (iii) a checkable assertion that needs to hold whenever the condition is met.
Records violate the dependency if the applicability condition holds but the assertion does not.
We denote a set of semantic dependencies by $\mathcal{C}$.
\end{definition}

\begin{example}
\label{ex:semantic_dependencies}
In a shopping database, the functional dependency $\{\texttt{quantity},\texttt{unit\_price}\}\rightarrow\texttt{amount}$ requires items with the same quantity and unit price to have the same amount.
The arithmetic rule $\texttt{amount}=\texttt{quantity}\times\texttt{unit\_price}$ specifies how that amount is computed.
Across tables, an order's \texttt{subtotal} equals the sum of the amounts of all its linked items.
A conditional rule may further require $\texttt{tax}=0$ for tax-exempt orders.
These semantic dependencies express requirements over individual records, related record groups, and records satisfying specific conditions.
\end{example}

\subsection{LM-based Data Generation}
\label{sec:prelim_llm}

\noindent\textbf{Autoregressive generation.}
LMs, including the GPT~\cite{gpt5} and Qwen~\cite{qwen3} families, acquire broad world knowledge and strong reasoning and generation capabilities through large-scale
text pretraining. 
% \textcolor{red}{Their autoregressive formulation supports distribution modeling
% and data generation.}
Given a conditioning input $c$ and a target sequence
$y=(y_1,\ldots,y_T)$, the conditional probability is
factorized as
\[
p_{\vartheta}(y\mid c)
=
\prod_{t=1}^{T}
p_{\vartheta}(y_t\mid c,y_{<t}),
\]
where $\vartheta$ denotes the model parameters and
$y_{<t}$ contains the preceding tokens.
New sequences are generated by successively sampling
tokens from these conditional distributions.
Existing methods, including GReaT for single-table
synthesis and REaLTabFormer for multi-table synthesis,
apply this autoregressive formulation to database
synthesis by serializing records as token sequences.

\noindent\textbf{Fine-tuning and soft prompts.}
% \textcolor{red}{Fine-tuning adapts a pretrained LM to a target task using task-specific data.}
In supervised fine-tuning (SFT)~\cite{sft}, the model maximizes
$p_{\vartheta}(y\mid c)$ over conditioning--target pairs $(c,y)$.
Target tokens are predicted given their observed predecessors, and the
sequence loss is used to optimize all or a subset of model parameters.

Soft prompts~\cite{softprompt} condition an LM through continuous
vectors in its token embedding space.
A soft prompt
$\mathbf{P}\in\mathbb{R}^{M\times d_{\mathrm{LLM}}}$
contains $M$ such vectors, where $d_{\mathrm{LLM}}$
is the token embedding dimension.
Together with an instruction $q$, it forms
the conditioning prefix $
\mathbf{H}^{0}
=
[\mathbf{P};\mathrm{Emb}(q)],
$
where $\mathrm{Emb}(q)$ denotes the instruction's
token embeddings and $[\cdot;\cdot]$ denotes
concatenation along the sequence dimension.
In conventional soft prompt tuning, the prompt vectors
are optimized while the pretrained LM remains frozen.

Building on these preliminaries, we present \our{}'s two stages:
Graph-to-Prompt Tuning (Section~\ref{sec:tuning}), which jointly optimizes
the SLM and a graph-to-prompt mapping through autoregressive reconstruction
of mixed-type record attributes, and Collaborative SLM--LLM Decoding (Section~\ref{sec:decoding}),
which synthesizes records and reduces semantic inconsistencies.

%% file: tuning.tex
% Section 3. Graph-to-Prompt Tuning.
\section{Graph-to-Prompt Tuning}
\label{sec:tuning}
\label{sec:overall_framework}

\begin{figure*}[t]
    \centering
    \includegraphics[width=0.96\textwidth]{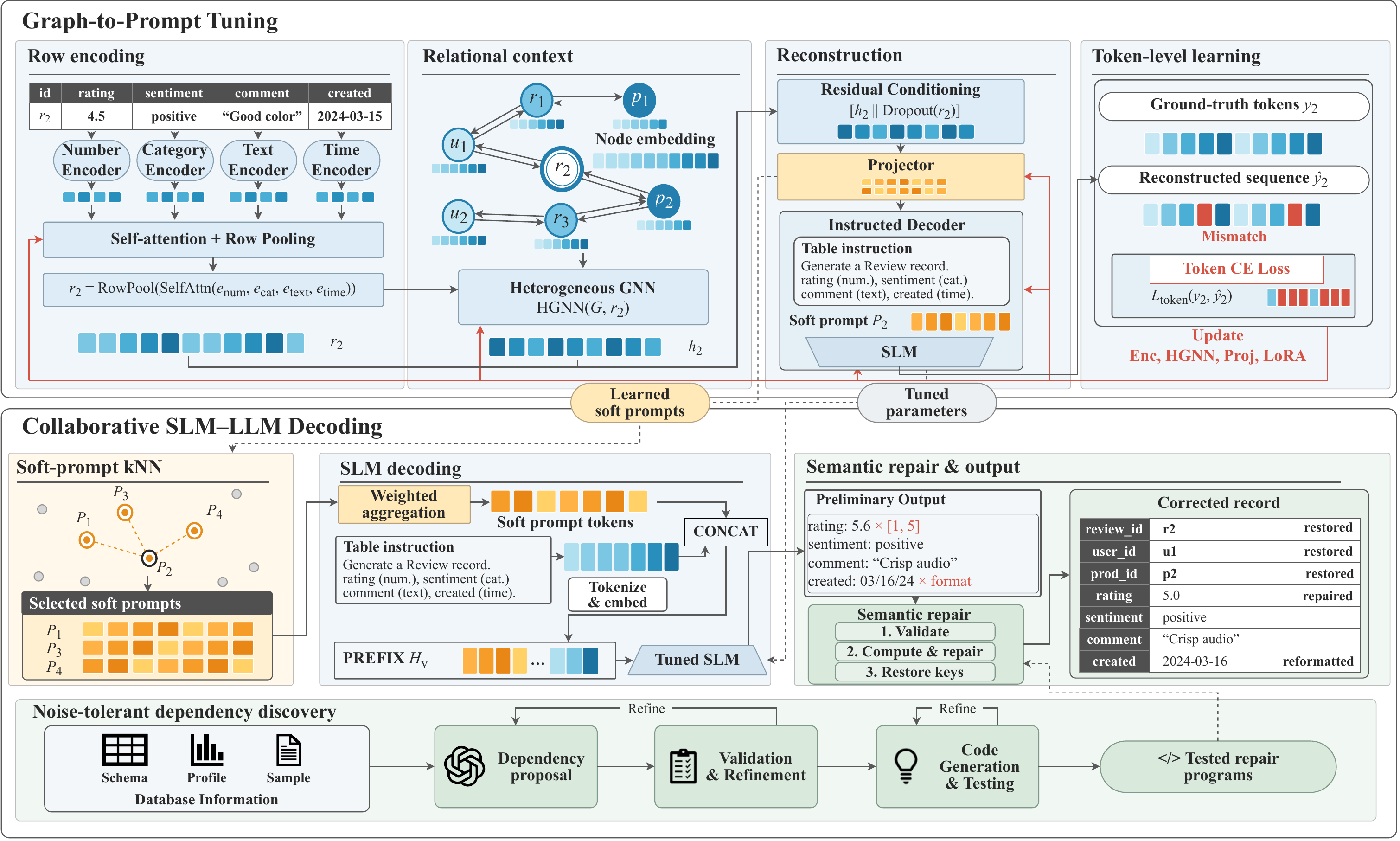}
    \caption{Overview of \our{}, consisting of Graph-to-Prompt Tuning and Collaborative SLM--LLM Decoding.}
    \label{fig:overview}
\end{figure*}

\subsection{Overview}
\label{sec:overview}

Given a source relational database, realistic synthesis of
mixed-type records requires modeling both the contents of individual records and their dependencies through foreign-key links.
To this end, \our{} jointly learns a soft-prompt representation space for relational records and tunes an SLM to model attribute distributions conditioned on the resulting prompts.
As illustrated in Figure~\ref{fig:overview}, \our{} first constructs an attributed heterogeneous graph and encodes each record's mixed-type attributes into a row embedding.
A heterogeneous GNN then aggregates relational context, which is combined with a masked row embedding and projected into soft prompts.
The SLM reconstructs record attributes conditioned on these prompts, providing a shared reconstruction objective for learning the mapping and adapting the SLM.

\subsection{Attributed Heterogeneous Graph Building}
\label{sec:ahg}

\noindent\textbf{Motivation.}
An attributed heterogeneous graph provides a natural
representation of relational databases, preserving
individual records and their foreign-key connections
in a unified structure. Record nodes retain attribute
information, while typed edges distinguish different
relationships and enable context aggregation across
connected records. Within each record, mixed-type
attributes carry different forms of information.
We therefore use type-aware encoding to preserve
these characteristics and incorporate column
semantics to identify what each value represents.

\noindent\textbf{Graph construction.}
Given a relational database $\mathcal{D}=(\mathcal{R},\mathcal{L})$, we construct an attributed heterogeneous graph $\mathcal{G}=(\mathcal{V},\mathcal{E},\phi,\psi,\mathbf{X})$ to preserve its native relational structure.
Here, $\mathcal{V}$ and $\mathcal{E}$ denote the node and edge sets, $\phi:\mathcal{V}\rightarrow\mathcal{R}$ maps each node to its table type, $\psi:\mathcal{E}\rightarrow\mathcal{L}\cup\mathcal{L}^{-1}$ maps each edge to its relation type, and $\mathbf{X}$ contains the non-key attributes.
The set $\mathcal{L}^{-1}$ contains the reverse of each foreign-key relationship.

For each row $r_i^{(a)}$ in table $R^{(a)}$, we create a node $v_i^{(a)}\in\mathcal{V}$ with node type $\phi(v_i^{(a)})=R^{(a)}$.
For each relationship $\rho\in\mathcal{L}$ from table $R^{(a)}$ to table $R^{(b)}$, if row $r_i^{(a)}$ references row $r_j^{(b)}$ through $\rho$, we add two directed edges:
\[
e_{ij}^{\rho}=(v_i^{(a)},v_j^{(b)},\rho),\qquad
e_{ji}^{\rho^{-1}}=(v_j^{(b)},v_i^{(a)},\rho^{-1}).
\]
Their edge types are defined as
\[
\psi(e_{ij}^{\rho})=\rho,\qquad
\psi(e_{ji}^{\rho^{-1}})=\rho^{-1}.
\]
These edge types distinguish messages from parent to child records and from child to parent records.

\noindent\textbf{Type-aware row encoding.}
Each record node is initialized with a row embedding of its non-key attributes.
For node $v_i^{(a)}$ corresponding to row $r_i^{(a)}$ in table $R^{(a)}$, we denote these $d_a$ attributes as $\mathbf{x}_i^{(a)}=\{x_{i1}^{(a)},\cdots,x_{id_a}^{(a)}\}$.
Their column names are $\{c_1^{(a)},\cdots,c_{d_a}^{(a)}\}$, with types $\tau(c_j^{(a)})\in\{\mathrm{num},\mathrm{cat},\mathrm{time},\mathrm{text}\}$.
We encode each attribute value according to its column type:
\[
\mathbf{u}_{ij}^{(a)}
=\mathrm{Enc}_{\tau(c_j^{(a)})}(x_{ij}^{(a)}).
\]
Numerical values are normalized and encoded with Fourier features, while categorical values use a vocabulary embedding table.
Temporal values are decomposed into calendar and linear-time features.
For textual values, a frozen text encoder encodes the column name and cell value together as ``table.column: value''.

For non-textual attributes, the value embedding alone is insufficient because the same value can have different meanings under different column names.
Therefore, \our{} further encodes each column name using the frozen text encoder:
\[
\mathbf{s}_{j}^{(a)}
=\mathrm{TextEnc}(\texttt{``}R^{(a)}.c_j^{(a)}\texttt{''}).
\]
The value embedding $\mathbf{u}_{ij}^{(a)}$ and the column-name embedding $\mathbf{s}_{j}^{(a)}$ are then projected into a shared embedding space and combined by a gated fusion module:
\[
\mathbf{e}_{ij}^{(a)}
=\mathbf{g}_{ij}^{(a)}\odot\bar{\mathbf{u}}_{ij}^{(a)}
+(1-\mathbf{g}_{ij}^{(a)})\odot\bar{\mathbf{s}}_j^{(a)},
\]
where $\bar{\mathbf{u}}_{ij}^{(a)}$ and $\bar{\mathbf{s}}_j^{(a)}$ are the projected value and column-name embeddings, and $\mathbf{g}_{ij}^{(a)}$ is a learnable gate.
% Column-wise scaling is absorbed into the projected embeddings.
Each fused attribute embedding captures both the value and its column meaning.
For textual attributes, we use the projected text embedding as $\mathbf{e}_{ij}^{(a)}$, since it already encodes the column name and cell value.

To capture dependencies among attributes within the same row, we apply a lightweight multi-head self-attention block over their embeddings.
The refined embeddings are mean-pooled and projected into a row embedding:
\[
\mathbf{r}_{v_i^{(a)}}
=\mathbf{W}_o\cdot\mathrm{MeanPool}\left(
\mathrm{SelfAttn}
\left(\{\mathbf{e}_{i1}^{(a)},\cdots,\mathbf{e}_{id_a}^{(a)}\}\right)
\right).
\]
The resulting row embedding $\mathbf{r}_{v_i^{(a)}}$ summarizes the mixed-type attributes of row $r_i^{(a)}$ and initializes the node for message passing.
% SelfAttn abstracts residual/normalization. Key-only tables use a learned
% empty-row embedding and provide context without being reconstruction targets.

\subsection{Graph-to-Prompt Mapping}
\label{sec:softprompt}

\noindent\textbf{Motivation.}
With the attributed heterogeneous graph in place, we need to translate its attribute and relational information into a form suitable for conditioning the SLM. We therefore adopt soft prompts as a learnable interface to the SLM's token embedding space, incorporating both relational context aggregated from connected records and the target record's own row embedding.

\noindent\textbf{Prompt information preparation.}
Given the graph $\mathcal{G}$ and row embeddings $\mathbf{r}_v$, a heterogeneous GNN aggregates information from related records into a context embedding.
Starting from $\mathbf{h}_v^{(0)}=\mathbf{r}_v$, each layer updates node $v$ by message passing:
\[
\mathbf{h}_v^{(g+1)}
=\sigma\left(
\mathbf{W}_{\phi(v)}^{(g)}\mathbf{h}_v^{(g)}
+\sum_{\rho\in(\mathcal{L}\cup\mathcal{L}^{-1})}
\operatorname{AGG}_{u\in\mathcal{N}_{\rho}(v)}
\mathbf{W}_{\rho}^{(g)}\mathbf{h}_u^{(g)}
\right),
\]
where $\mathcal{N}_{\rho}(v)$ denotes the neighbors of $v$ connected by relation type $\rho$.
The matrices $\mathbf{W}_{\phi(v)}^{(g)}$ and $\mathbf{W}_{\rho}^{(g)}$ transform messages according to node and relation types, while $\operatorname{AGG}$ and $\sigma$ denote neighborhood aggregation and a nonlinear transformation.
After $L$ layers, the context embedding $\mathbf{h}_v=\mathbf{h}_v^{(L)}$ contains information from parent records, child records, and multi-hop relational neighborhoods.
% This is the original abstract message-passing form: W_phi(v) summarizes
% relation-specific self transforms, and sigma includes normalization/activation.

Repeated message passing can make record-specific information less distinguishable in the context embedding.
\our{} therefore retains a direct row residual and randomly masks it to vary the reconstruction condition:
\[
\widetilde{\mathbf{r}}_v
=\frac{\mathbf{m}_v\odot\mathbf{r}_v}{1-p},
\qquad
\mathbf{m}_v\sim\mathrm{Bernoulli}(1-p),
\]
where $p$ is the dropout ratio and $\mathbf{m}_v$ is a random mask.
% The GNN uses the unmasked row embeddings; dropout applies to the residual.
We concatenate the context embedding with this masked row embedding:
\[
\mathbf{z}_v=[\mathbf{h}_v\Vert\widetilde{\mathbf{r}}_v].
\]

\noindent\textbf{Soft prompt construction.}
A learnable projector maps the concatenated embedding $\mathbf{z}_v$ into a soft prompt in the SLM's token embedding space:
\[
\mathbf{P}_v=\mathrm{Projector}(\mathbf{z}_v)
\in\mathbb{R}^{M\times d_{\mathrm{SLM}}},
\]
where $M$ is the number of soft prompt vectors and $d_{\mathrm{SLM}}$ is the token embedding dimension of the SLM decoder.

For each target table $R^{(a)}$, \our{} constructs a textual instruction $q_{\phi(v)}$ specifying the table name, non-key columns, their types, and the output format.
The instruction follows a canonical template:
\[
\begin{aligned}
q_{\phi(v)}=\;&\texttt{Generate attributes for }\phi(v)\texttt{.}\\
&\texttt{Columns: }c_1^{(a)}\texttt{(}\tau(c_1^{(a)})\texttt{), }\cdots\texttt{.}\\
&\texttt{Format: }c_1^{(a)}\texttt{: value;; }\cdots\texttt{.}
\end{aligned}
\]
Column types use descriptors such as \texttt{num}, \texttt{cat}, \texttt{time}, or \texttt{text}.
We concatenate the soft prompt with the instruction's token embeddings to form the decoder input:
\[
\mathbf{H}_v^0=[\mathbf{P}_v;\mathrm{Emb}(q_{\phi(v)})].
\]
The instruction specifies which attributes to reconstruct, while the soft prompt supplies their row information and relational context.

\subsection{Tuning Procedure}
\label{sec:training}

\noindent\textbf{Motivation.}
To learn mixed-type attribute distributions conditioned on the soft prompts constructed above, we jointly tune the graph-to-prompt mapping and the SLM through autoregressive attribute reconstruction.
The reconstruction task provides a shared learning signal, guiding the mapping to produce informative prompts and the SLM to generate attributes conditioned on them.

\noindent\textbf{Training objective.}
For supervision, we serialize each record's non-key attributes into a target sequence
$\mathbf{y}_v=(y_{v,1},\cdots,y_{v,T_v})$.
Deterministically computable attributes are omitted from these targets and restored after decoding by the tested programs in Section~\ref{sec:constraint_engine}.
Specifically, for numerical columns that are difficult for token-based prediction, especially those with high-cardinality or continuous values, we use quantile-based textual representations.
For categorical columns with long labels, we use short codes such as \texttt{C1}.
For example, a serialized record may be ``\texttt{price(quantile): 0.7321;;}
\texttt{room\_type: C1;;}
\texttt{review: near the metro;;}''.
Conditioned on the input embeddings $\mathbf{H}_v^0$ constructed in Section~\ref{sec:softprompt}, the SLM reconstructs this sequence autoregressively under teacher forcing.
The tuning objective is to minimize the normalized autoregressive loss:
\[
\mathcal{J}(\Theta)
=
-\frac{1}{\sum_{v\in\mathcal{V}_{tr}} T_v}
\sum_{v\in\mathcal{V}_{tr}}
\sum_{t=1}^{T_v}
\log p_{\Theta}
\left(y_{v,t}\mid\mathbf{H}_v^0,y_{v,<t}\right).
\]
Here, $\mathcal{V}_{tr}$ is the set of training target nodes, and $\Theta$ contains the trainable parameters of the type-aware row encoder, heterogeneous GNN, soft-prompt projector, and adapted SLM.
We adapt the SLM using LoRA~\cite{lora} while keeping the separate text encoder frozen.

\noindent\textbf{Training pipeline.}
Algorithm~\ref{alg:training} summarizes the tuning procedure.
In each iteration, \our{} samples $B_{\mathrm{tr}}$ target record nodes with non-key attributes and expands their $L$-hop neighborhoods to provide relational context.
Only the target nodes contribute to the training objective, while the sampled neighbor nodes supply information for message passing.
\our{} constructs soft prompts for the target nodes and updates $\Theta$ using their autoregressive loss.
The tuned model, parameterized by $\Theta$, is used for attribute generation in the subsequent decoding phase.

\begin{algorithm}[t]
\caption{Graph-to-Prompt Tuning of \our{}}
\label{alg:training}
\begin{minipage}{\hsize}
\begin{algorithmic}[1]
\REQUIRE Database $\mathcal{D}$, seed $s_1$, steps $S_{\mathrm{tr}}$, target batch size $B_{\mathrm{tr}}$, depth $L$
\ENSURE Tuned parameters $\Theta$
\STATE Build attributed graph $\mathcal{G}$ from $\mathcal{D}$
\STATE Cache frozen text features and serialization mappings
\STATE Serialize non-key attributes as labels $\{\mathbf{y}_v\}$
\STATE Initialize $\Theta$ and set random seed $s_1$
\FOR{$s=1,\ldots,S_{\mathrm{tr}}$}
    \STATE $\mathcal{V}_b\leftarrow\operatorname{SampleTargets}(\mathcal{G},B_{\mathrm{tr}})$
    \STATE $\mathcal{G}_b\leftarrow\operatorname{KHopSubgraph}(\mathcal{G},\mathcal{V}_b,L)$
    \STATE Encode rows in $\mathcal{G}_b$ to obtain $\{\mathbf{r}_v\}$
    \STATE Propagate context on $\mathcal{G}_b$ to obtain $\{\mathbf{h}_v\}$
    \FOR{$v\in\mathcal{V}_b$}
        \STATE $\widetilde{\mathbf{r}}_v\leftarrow\operatorname{Dropout}(\mathbf{r}_v)$
        \STATE $\mathbf{P}_v\leftarrow\operatorname{Projector}_{\Theta}([\mathbf{h}_v\Vert\widetilde{\mathbf{r}}_v])$
    \ENDFOR
    \STATE Compute normalized autoregressive loss $\mathcal{J}$ on $\mathcal{V}_b$
    \STATE $\Theta\leftarrow\operatorname{OptimizerStep}(\Theta,\nabla_{\Theta}\mathcal{J})$
\ENDFOR
\RETURN $\Theta$
\end{algorithmic}
\end{minipage}
\end{algorithm}

%% file: decoding.tex
% Sections 4 and 5: collaborative decoding and shared complexity analysis.
\section{Collaborative SLM--LLM Decoding}
\label{sec:decoding}
\label{sec:training_generation}

\subsection{Overview}
\label{sec:decoding_overview}

The graph-to-prompt mapping and tuned SLM from Section~\ref{sec:tuning} provide the basis for generating mixed-type records conditioned on relational context. However, probabilistic
decoding from learned attribute distributions does not guarantee that generated values satisfy semantic
dependencies. To address this, \our{} combines SLM-based record generation with an LLM-driven semantic dependency engine.
As illustrated in Figure~\ref{fig:overview}, the tuned SLM generates mixed-type attributes from soft prompts, while an LLM discovers semantic dependencies and refines them using independent validation feedback.
The validated dependencies are translated into tested repair programs, which are applied to the generated database to correct semantic inconsistencies.

\subsection{SLM-Based Record Decoding}
\label{sec:slm_decoding}

\noindent\textbf{Motivation.}
%Graph-to-Prompt Tuning learns a soft-prompt representation space for relational records and tunes the SLM to model attribute distributions conditioned on these prompts.
%By varying the row masks and combining nearby prompts from the same table in the learned soft-prompt space, \our{} constructs new conditioning inputs for synthesis.
While training reconstructs source records from their soft prompts, synthesis requires constructing new conditioning inputs rather than simply reusing those prompts. We therefore vary the row masks and aggregate nearby prompts from the same table in the learned soft-prompt space.
The resulting soft prompts are designed to introduce variation while maintaining semantic similarity, guiding the tuned SLM to generate synthetic records from the learned conditional distributions.

\noindent\textbf{Prompt preparation.}
Given the attributed heterogeneous graph of $\mathcal{D}$, \our{} uses the tuned row encoder and heterogeneous GNN to obtain row and context embeddings.
For each record, we mask its row embedding using an inference seed $s_2$ distinct from the tuning seed $s_1$, which varies the row information supplied for synthesis.
The context embedding is concatenated with the masked row embedding and projected into a soft prompt, following Section~\ref{sec:softprompt}.

We then select other records from the same table using $k$-nearest-neighbor (kNN) search in soft-prompt space to draw on semantically similar records.
Their prompts are combined with normalized inverse-distance weights to form the synthesis prompt $\mathbf{P}_v$.
% If no eligible neighbor exists, the original prompt is retained.
As in Section~\ref{sec:softprompt}, we concatenate $\mathbf{P}_v$ with the token embeddings of the table-specific instruction $q_{\phi(v)}$ to form the decoder input $\mathbf{H}_v^0=[\mathbf{P}_v;\mathrm{Emb}(q_{\phi(v)})]$.

\noindent\textbf{Attribute decoding.}
Conditioned on $\mathbf{H}_v^0$, the tuned SLM generates the non-key attributes of a new synthetic record $\widetilde{r}_v$.
The output sequence $\widetilde{\mathbf{y}}_v$ has autoregressive likelihood:
\[
p_{\Theta}(\widetilde{\mathbf{y}}_v \mid \mathbf{H}_v^0)
=
\prod_{t=1}^{|\widetilde{\mathbf{y}}_v|}
p_{\Theta}
\left(
\widetilde{y}_{v,t}
\mid
\mathbf{H}_v^0, \widetilde{y}_{v,<t}
\right).
\]
Here, $\Theta$ denotes the tuned generator parameters.
The output follows the column-value format specified by the instruction:
\[
\begin{aligned}
&
\texttt{column}_1\texttt{: value}_1\texttt{;; }
\texttt{column}_2\texttt{: value}_2\texttt{;; }\\
&\cdots\;
\texttt{column}_{d_a}\texttt{: value}_{d_a}\texttt{;;}
\end{aligned}
\]

We then parse each generated sequence into typed values using the serialization mappings introduced in Section~\ref{sec:training}.
During parsing, generated quantile values are mapped back to the source numeric scale, and categorical codes are converted to their original labels.
We place the resulting records in the synthetic database with the source schema, table sizes, and key assignments.

\subsection{LLM-Driven Semantic Dependency Engine}
\label{sec:constraint_engine}

\noindent\textbf{Motivation.}
The probabilistic nature of SLM decoding can lead to
violations of semantic dependencies among generated values.
Repairing these violations requires identifying dependencies
that are often implicit in the data and difficult to infer
from the schema alone.
A naive approach is to ask an LLM to enumerate all
dependencies directly from the database, but the resulting
dependency set can be incomplete and inaccurate.
Noise in the source database further complicates discovery, as even valid
dependencies can be violated occasionally.
\our{} therefore combines LLM-guided proposals with
iterative validation and refinement for noise-tolerant
semantic dependency discovery.

\noindent\textbf{Discovery and refinement.}
To identify candidate semantic dependencies, \our{} provides
the LLM with the schema, data profiles, and relational
examples from the source database.
We instruct the LLM to combine this context with its world
knowledge to identify determining and dependent attributes
for \emph{functional dependencies}, formulate equations
for \emph{arithmetic relationships}, and infer semantically
justified value domains, conditional implications, and
event consistency requirements for \emph{domain-specific
rules}, both within and across tables.
The LLM returns at most $Q$ candidates, each specifying a
checkable assertion, its scope, and an applicability
condition.
The discovery instructions are summarized in
Box~\ref{box:dependency_discovery_prompt}. 
The full prompt templates are provided in our source code.

\begin{promptbox}[float=t,label={box:dependency_discovery_prompt}]
\textbf{Input context.}
Schema, data profiles, and relational examples from the
source database.

\par\smallskip
\textbf{Discovery guidance.}
Combine the supplied data evidence with your world
knowledge to identify dependencies within and across tables:
\begin{itemize}[
    leftmargin=1.2em,
    topsep=3pt,
    itemsep=2pt,
    parsep=0pt,
    partopsep=0pt
]
    \item Functional dependencies: identify attributes
    determined by one or more others, including semantic
    hierarchies.
    \item Arithmetic relationships: formulate equations, comparisons.
    \item Domain-specific rules: identify specific value
    domains, conditional implications, and event consistency
    requirements.
\end{itemize}
Distinguish semantic requirements from frequent patterns;
allow plausible rare exceptions.

\par\smallskip
\textbf{Output.}
Return at most $Q$ candidates in JSON, each with an assertion,
a scope, and an applicability condition.
\end{promptbox}

However, these candidates can be overly strict or based on
incorrect semantic assumptions.
To identify valid dependencies while accommodating source
noise, \our{} therefore iteratively validates and refines
them over at most $R_{\mathrm{val}}$ rounds, allowing a small
tolerance $\epsilon$ for the true violation probability  of each dependency.
In each round, \our{} validates candidates on fresh samples
from the source database.
To account for sampling uncertainty, we accept a candidate
only if the upper confidence bound on its violation
probability is at most $\epsilon$.
Specifically, for a candidate $\gamma$, let $n_\gamma>0$
be the number of total evaluation units, $b_\gamma$
the number of violated units, and $p_\gamma$ the true
violation probability.
We model their violation indicators as i.i.d.\ Bernoulli
variables, giving
$b_\gamma\sim\mathrm{Binomial}(n_\gamma,p_\gamma)$. The one-sided Clopper--Pearson
upper confidence bound~\cite{ucp} $U_{\mathrm{CP}}$
at confidence level $1-\alpha$ satisfies
\[
\sum_{i=0}^{b_\gamma}\binom{n_\gamma}{i}
U_{\mathrm{CP}}^{\,i}
(1-U_{\mathrm{CP}})^{n_\gamma-i}
=\alpha.
\]
The acceptance criterion $U_{\mathrm{CP}}\leq\epsilon$
limits the probability of accepting a dependency with
$p_\gamma>\epsilon$ to at most $\alpha$ in each test.
To maintain this guarantee across iterative refinement,
we allocate a total error budget $\delta$ over at most
$Q$ candidates and $R_{\mathrm{val}}$ rounds, setting
$\alpha=\delta/(Q R_{\mathrm{val}})$.
A union bound then gives
\[
\Pr\!\left(
    \exists\gamma\in\mathcal{C}^{*}:
    p_\gamma>\epsilon
\right)
\leq Q R_{\mathrm{val}}\alpha=\delta,
\]
where $\mathcal{C}^{*}$ denotes the dependencies accepted
by this criterion.
Thus, with confidence at least $1-\delta$, all dependencies
in $\mathcal{C}^{*}$ have violation probabilities at most
$\epsilon$.
Accepted dependencies then remain fixed.
For the remaining candidates, the LLM uses validation
statistics, counterexamples, and satisfying instances to
revise their definitions, discard invalid ones, or retain
those requiring more evidence.
Revised candidates are retested, and the loop ends when
no candidates remain pending or the round budget is
exhausted.

The LLM then implements the validated dependencies as repair
programs, which are tested individually and jointly on
source and perturbed records for violation correction and
preservation of valid inputs.
During synthesis, these tested programs repair semantic
inconsistencies in the database generated by the SLM.

\subsection{Collaborative Decoding Procedure}
\label{sec:generation}

Algorithm~\ref{alg:generation} summarizes the collaborative decoding procedure of \our{}.
Given the source database $\mathcal{D}$, \our{} first obtains validated dependencies $\mathcal{C}^{*}$ and tested repair functions $\mathcal{F}$ through the semantic dependency engine and caches them for subsequent repair.
With the tuned parameters $\Theta$, the complete generation process is
\[
\begin{aligned}
\widetilde{\mathcal{D}}^{(0)}
&=\operatorname{Generate}_{\Theta}(\mathcal{D};s_2),\\
\widetilde{\mathcal{D}}
&=\operatorname{Execute}_{\mathcal{F},\mathcal{C}^{*}}
(\widetilde{\mathcal{D}}^{(0)}).
\end{aligned}
\]

Here, $s_2$ is the inference seed, and $\operatorname{Generate}_{\Theta}$ includes graph encoding, soft-prompt preparation, SLM decoding, and parsing the outputs into typed values.
The decoded non-key attributes are assembled into $\widetilde{\mathcal{D}}^{(0)}$ with the source schema, table sizes, and key assignments.
After all tables have been populated, $\operatorname{Execute}_{\mathcal{F},\mathcal{C}^{*}}$ applies the tested repair functions in the planned dependency order.
Interacting updates are checked together against affected dependencies, and changes that introduce new violations are rolled back.

\begin{algorithm}[t]
\caption{Collaborative SLM--LLM Decoding of \our{}}
\label{alg:generation}
\begin{minipage}{\hsize}
\begin{algorithmic}[1]
\REQUIRE Database $\mathcal{D}$, tuned parameters $\Theta$, seed $s_2$, neighborhood depth $L$
\ENSURE Synthetic relational database $\widetilde{\mathcal{D}}$
\STATE Load or construct $(\mathcal{C}^{*},\mathcal{F})\leftarrow\operatorname{DependencyEngine}(\mathcal{D})$
\STATE Reuse or build attributed graph $\mathcal{G}$ from $\mathcal{D}$
\STATE Initialize $\widetilde{\mathcal{D}}^{(0)}$ with source schema and keys
\STATE $\operatorname{SetSeed}(s_2)$
\FOR{$\mathcal{V}_b\in\operatorname{BatchAttributeTargets}(\mathcal{G})$}
    \STATE $\mathcal{G}_b\leftarrow\operatorname{KHopSubgraph}(\mathcal{G},\mathcal{V}_b,L)$
    \STATE Encode rows in $\mathcal{G}_b$ to obtain $\{\mathbf{r}_v\}$
    \STATE Propagate relational context on $\mathcal{G}_b$ to obtain $\{\mathbf{h}_v\}$
    \FOR{$v\in\mathcal{V}_b$}
        \STATE $\widetilde{\mathbf{r}}_v\leftarrow\operatorname{Dropout}(\mathbf{r}_v)$
        \STATE $\mathbf{P}_v\leftarrow\operatorname{Projector}_{\Theta}([\mathbf{h}_v\Vert\widetilde{\mathbf{r}}_v])$
    \ENDFOR
\ENDFOR
\STATE $\{\mathbf{P}_v\}\leftarrow\operatorname{PrepareInferenceInputs}(\{\mathbf{P}_v\},\mathcal{G})$
\FOR{$\mathcal{V}_b\in\operatorname{BatchAttributeTargets}(\mathcal{G})$}
    \FOR{$v\in\mathcal{V}_b$}
        \STATE $\widetilde{\mathbf{y}}_v\leftarrow\operatorname{Decode}_{\Theta}(\mathbf{P}_v,q_{\phi(v)})$
        \STATE $\widetilde{r}_v\leftarrow\operatorname{Parse}_{\phi(v)}(\widetilde{\mathbf{y}}_v)$
        \STATE Write non-key values of $\widetilde{r}_v$ into $\widetilde{\mathcal{D}}^{(0)}$
    \ENDFOR
\ENDFOR
\STATE $\widetilde{\mathcal{D}}\leftarrow\operatorname{Execute}_{\mathcal{F},\mathcal{C}^{*}}(\widetilde{\mathcal{D}}^{(0)})$
\STATE Audit $\widetilde{\mathcal{D}}$ under $\mathcal{C}^{*}$
\RETURN $\widetilde{\mathcal{D}}$
\end{algorithmic}
\end{minipage}
\end{algorithm}

\section{Complexity Analysis}
\label{sec:complexity}

We analyze the tuning and decoding costs of \our{} in Sections~\ref{sec:tuning} and~\ref{sec:decoding}, respectively.
Let $N=|\mathcal{V}|$ be the number of records, $d_{\max}$ the maximum number of non-key attributes per record, and $\ell$ the maximum decoder sequence length, including soft prompts, instructions, and attribute tokens.
The analysis assumes fixed neural widths and decoder architecture.
Preparation cost $T_{\mathrm{feat}}$ covers graph construction and frozen text encoding, which are reused during decoding.

\noindent\textbf{Tuning.}
During tuning, let $n_b$ and $e_b$ bound the numbers of sampled nodes and edges per batch of $B_{\mathrm{tr}}$ target records.
Row attention, $L$-layer graph propagation, and teacher-forced decoding cost $O(n_b d_{\max}^{2})$, $O(L(n_b+e_b))$, and $O(B_{\mathrm{tr}}\ell^{2})$, respectively.
Backpropagation preserves these asymptotic orders.
The overall cost of $S_{\mathrm{tr}}$ tuning steps, including preparation, is given by
\[
T_{\mathrm{feat}}+
O\!\left(S_{\mathrm{tr}}\left[
n_b d_{\max}^2+L(n_b+e_b)+B_{\mathrm{tr}}\ell^2
\right]\right).
\]

\noindent\textbf{Decoding and repair.}
During decoding, let $N_{\Sigma}$ and $E_{\Sigma}$ count accumulated node and edge visits across sampled subgraphs, including repeated neighborhoods.
Let $T_{\mathrm{DE}}$ cover semantic dependency discovery, validation, and repair-program preparation and testing.
With source features cached, the first synthesis run costs
\[
\begin{aligned}
T_{\mathrm{DE}}+O\!\left(
N_{\Sigma}d_{\max}^2+L(N_{\Sigma}+E_{\Sigma})+N\ell^2
\right)
{}+T_{\mathrm{kNN}}+T_{\mathrm{repair}}.
\end{aligned}
\]
The neural terms cover row encoding, graph propagation, and autoregressive decoding with a key/value cache.
Here, $T_{\mathrm{kNN}}$ includes neighbor search and prompt combination, and $T_{\mathrm{repair}}$ includes program execution and auditing.
The current exact search compares prompts within each table, giving $T_{\mathrm{kNN}}=O(Md_{\mathrm{SLM}}\sum_a n_a^2)$ for fixed $k$, where $n_a$ is the size of table $R^{(a)}$.
Chunking the search reduces peak memory but does not remove this quadratic time term.

The cost of semantic dependency validation and repair depends on the scopes of the retained dependencies.
Bounded-size row checks and indexed key-link checks take $O(|\mathcal{C}^*|N)$, whereas ordered checks may require $O(|\mathcal{C}^*|N\log N)$.
General joins, repeated checks, and joint repair solvers depend on intermediate sizes and execution plans and remain within $T_{\mathrm{DE}}$ and $T_{\mathrm{repair}}$.
Subsequent runs reuse validated dependencies and tested programs, avoiding $T_{\mathrm{DE}}$ when these artifacts remain applicable.
The prompt bank uses $O(NMd_{\mathrm{SLM}})$ space; for decoding batches of $B$ records, the key/value cache uses $O(B\ell)$ additional space under the fixed-architecture assumption, besides graph storage, cached features, search buffers, and repair intermediates.

%% file: experiment.tex
\newcommand{\best}[1]{\textbf{#1}}
\newcommand{\second}[1]{\underline{#1}}
\newcommand{\NA}{--}

% =========================
% experiment_evaluation.tex
% =========================
% TODO (global): once all baseline numbers are final, mark the best and
% second-best results in each table with \best{} and \second{}, and fill
% the "Improv." columns.

\section{Experimental Evaluation}
\label{sec:experiments}

\subsection{Experiment Setup}
\label{sec:exp_setup}

\noindent\textbf{Dataset description.}
The evaluation uses seven real-world relational databases, whose profiles are summarized in Table~\ref{tab:datasets}. 
%For comparison with existing relational database generators, we use four databases without text: Biodegradability (Biodeg.), Walmart, and F1 (w/o text) following SyntheRela~\cite{syntherela}, and HM sourced from RelBench~\cite{relbench}.
%Following SyntheRela~\cite{syntherela}, we select four databases without text, i.e., Biodeg, Walmart, F1 (w/o text) and HM, to enable direct comparison with existing relational database generators. 
%To evaluate synthesis with textual attributes, a setting rarely covered by prior benchmarks, we further include three text-rich databases: F1 (w/ text) built from RelBench~\cite{relbench}, with text fields such as driver and circuit names; ML-small sourced from the MovieLens dataset~\cite{ml}, containing movie titles and user-generated tags; and Chicago constructed from the Chicago Open Data Portal~\cite{chicago_crimes,chicago_iucr}, including crime and location descriptions. These databases span diverse schema sizes, relationship patterns, and attribute types.
For comparison with existing relational database generators, we use four text-free databases: Biodegradability (Biodeg.), Walmart, and F1 (w/o text) following SyntheRela~\cite{syntherela}, and HM from RelBench~\cite{relbench}. To evaluate synthesis with textual attributes,
we further include three text-rich databases: F1 (w/ text) built from RelBench~\cite{relbench}, with driver and circuit names; ML-small from MovieLens~\cite{ml}, containing movie titles and user-generated tags; and Chicago from the Chicago Open Data~\cite{chicago_crimes,chicago_iucr}, including crime and location descriptions.

\begin{table}[t]
  \centering
  \captionsetup{skip=4pt}
  \caption{
Profiles of the evaluation databases.
$|\mathcal{R}|$, $N$, $d$, and $|\mathcal{L}|$ denote the numbers
of tables, rows, columns, and relationships, respectively.
MC denotes multiple children, and MC\&P denotes multiple children
and parents.
N, C, and T denote numerical, categorical, and text attributes.
}
  \label{tab:datasets}
  \small
  \setlength{\tabcolsep}{4pt}
  \renewcommand{\arraystretch}{1.10}
  \resizebox{\columnwidth}{!}{%
  \begin{tabular}{|c|c|c|c|c|c|c|}
    \hline
    \rowcolor[gray]{0.8}
    \textbf{Datasets} & $\boldsymbol{|\mathcal{R}|}$ & $\boldsymbol{N}$ & $\boldsymbol{d}$ & $\boldsymbol{|\mathcal{L}|}$ & \textbf{Hierarchy} & \textbf{Types} \\
    \hline
    Walmart         & 3 & 12,075  & 20 & 2  & MC    & N, C    \\
    \hline
    Biodeg.         & 5 & 21,895  & 16 & 5  & MC\&P & N, C    \\
    \hline
    F1 (w/o text) & 9 & 97,606  & 50 & 13 & MC\&P & N, C    \\
    \hline
    HM              & 3 & 97,917  & 24 & 2  & MC\&P & N, C    \\
    \hline
    Chicago         & 3 & 50,335  & 26 & 2  & MC\&P & N, C, T \\
    \hline
    F1 (w/ text)   & 9 & 97,606  & 50 & 13 & MC\&P & N, C, T \\
    \hline
    ML-small          & 4 & 102,806 & 13 & 4  & MC\&P & N, T    \\
    \hline
  \end{tabular}%
  }
\end{table}

\noindent\textbf{Baseline methods.}
We compare \our{} with seven representative baselines: 1) SDV~\cite{sdv}, which uses statistical modeling and copulas for synthesis; 2) RC-TGAN~\cite{rctgan}, which extends GAN-based generation to relational databases; 3) TabularARGN (TabARGN)~\cite{argn}, which uses autoregressive modeling to synthesize relational databases; 4) GRDM~\cite{grdm}, a graph-based diffusion model for relational database synthesis, selected as a published representative of this family over the methodologically similar RelDiff~\cite{reldiff}, which is available as a preprint; and 5) REaLTabFormer (REaLTabF.)~\cite{realtabformer}, which uses an LLM-based model for parent-child table generation.  Since existing relational database generators do not natively support textual attributes, we additionally implement two multi-table extensions of the single-table synthesizer GReaT~\cite{great}, enabling a direct comparison with \our{} on text-rich relational databases: 6) GReaT-Cond synthesizes tables in parent-to-child order, conditioning each child table on its corresponding parent records; and 7) GReaT-Joint joins each leaf table with its parents and ancestors through many-to-one relationships, applies GReaT-style generation to the resulting wide records, and splits them back into separate tables.

\noindent\textbf{Metrics.}
To evaluate synthesis quality, we follow the evaluation protocol of SyntheRela~\cite{syntherela}. We further adopt and extend the constraint violation metrics used by Constrained Deep Generative Models (C-DGMs)~\cite{cvr1} to measure semantic consistency in synthetic databases.
Specifically, we report five metrics: 1) classifier two-sample test with aggregation (C2ST-Agg), which uses a classifier to assess how distinguishable synthetic records are from real records based on relationally aggregated features; 2) Intra-table similarity, which measures how well pairwise relationships between columns within each table are preserved; 3) Inter-table similarity, which measures how well pairwise relationships between columns across directly related tables are preserved; 4) constraint violation rate (CVR), for which we follow C-DGMs~\cite{cvr1} by constructing a human-verified reference set of semantic dependencies and reporting the fraction of synthetic records violating at least one dependency; and 5) Utility, which measures downstream performance under the train-on-synthetic-test-on-real (TSTR) protocol, using AUC for classification and MAE for regression. 
For text-free databases, we retain the original fidelity and utility metric definitions in SyntheRela. For text-rich databases, we incorporate text-derived features into C2ST-Agg and text embeddings into downstream predictors, and extend intra- and inter-table similarity using embedding correlations and category-conditional text distribution similarity. CVR follows the same record-level definition in both settings.

\noindent\textbf{Implementation Details.}
For \our{}, we use DeepSeek-V4-Pro for the semantic dependency engine, all-MiniLM-L6-v2 for text embeddings, and Qwen3-0.6B as the decoder. The type-aware encoder outputs 384-dimensional row representations. For prompt information preparation, we use a two-layer heterogeneous GraphSAGE~\cite{sage} with 512 hidden dimensions to encode relational context, set the residual mask ratio to 0.4, and use a two-layer MLP projector to produce one soft prompt token. We train each model for at most 20K steps with batch size 8, learning rate $1\times10^{-4}$. During synthesis, we replace each soft prompt with the aggregation of its 3 nearest same-table neighbors and sample from the best checkpoint with temperature 0.8. For the semantic dependency engine, we propose at most 30 candidate dependencies and perform up to 5 rounds of iterative validation and revision. We set the violation tolerance to 0.1 and the total error budget to 0.1. For all baselines, we use their official implementations and default hyperparameters, with graph-based methods keeping the original relational graph fixed during generation. All methods are run with three random seeds on NVIDIA H800 GPUs with 80GB memory, and we report the mean and standard deviation.
The source code is available at \url{https://github.com/scsystem1/REMIX}.

\begin{table*}[t]
\centering
\caption{
Synthesis quality on text-free relational databases.
Results are mean $\pm$ standard deviation over three runs.
All metrics except TSTR are reported in \%.
Arrows indicate the direction of improvement.
Shaded bold and underlined values mark the best and second-best
results, respectively.
Blue rows report mean relative gains of \our{} over each baseline
across available datasets.
Dashes indicate unavailable results.
}
\label{tab:quality_notext}
\small
\setlength{\tabcolsep}{3.8pt}
\renewcommand{\arraystretch}{0.95}
\resizebox{\textwidth}{!}{%
\begin{tabular}{
@{}llccccccc
@{\hspace{4pt}\vrule width 0.45pt\hspace{4pt}}
c@{}
}
\toprule
Metrics & Datasets
& SDV
& RC-TGAN
& GRDM
& TabARGN
& REaLTabF.
& GReaT-Cond
& GReaT-Joint
& \our{} \\
\midrule

\multirow{4}{*}{C2ST-Agg ($\downarrow$)}
& Walmart
& $87.41_{\pm 2.31}$
& $93.70_{\pm 0.64}$
& \sotacell{65.56_{\pm 5.09}}
& $92.22_{\pm 2.94}$
& $91.85_{\pm 3.57}$
& $94.07_{\pm 0.64}$
& $98.89_{\pm 0.00}$
& $\underline{71.85_{\pm 3.57}}$ \\

& Biodeg.
& $87.83_{\pm 0.29}$
& $74.64_{\pm 0.19}$
& $\underline{68.43_{\pm 0.59}}$
& $84.96_{\pm 0.54}$
& --
& $85.57_{\pm 0.07}$
& $82.72_{\pm 0.10}$
& \sotacell{57.67_{\pm 0.45}} \\

& F1 (w/o text)
& $94.90_{\pm 0.35}$
& $92.63_{\pm 1.23}$
& $\underline{81.05_{\pm 0.92}}$
& $95.05_{\pm 0.54}$
& --
& $89.78_{\pm 0.69}$
& $90.71_{\pm 0.42}$
& \sotacell{64.97_{\pm 0.50}} \\

& HM
& $88.01_{\pm 1.18}$
& $87.00_{\pm 0.57}$
& $82.65_{\pm 0.46}$
& $\underline{81.53_{\pm 0.25}}$
& $86.56_{\pm 0.23}$
& $85.58_{\pm 0.15}$
& $90.53_{\pm 0.06}$
& \sotacell{79.79_{\pm 0.29}} \\

\addlinespace[2pt]
\multicolumn{2}{l}{
  \cellcolor{gainbg}\textit{Avg. Improv. of \our{}}
}
& \gaincell{+23.25\%}  % SDV
& \gaincell{+21.05\%}  % RC-TGAN
& \gaincell{+7.36\%}   % GRDM
& \gaincell{+22.00\%}  % TabARGN
& \gaincell{+14.80\%}  % REaLTabF.
& \gaincell{+22.66\%}  % GReaT-Cond
& \gaincell{+24.47\%}  % GReaT-Joint
& \cellcolor{gainbg}-- \\

\midrule

\multirow{4}{*}{Intra-table ($\uparrow$)}
& Walmart
& $87.08_{\pm 0.00}$
& $86.88_{\pm 1.10}$
& \sotacell{96.29_{\pm 0.34}}
& $86.86_{\pm 0.56}$
& $87.89_{\pm 0.80}$
& $91.94_{\pm 0.76}$
& $90.56_{\pm 0.37}$
& $\underline{93.59_{\pm 0.47}}$ \\

& Biodeg.
& $\underline{96.46_{\pm 0.00}}$
& $85.01_{\pm 3.58}$
& $94.79_{\pm 1.39}$
& $72.54_{\pm 0.34}$
& --
& $94.73_{\pm 1.05}$
& $95.25_{\pm 1.70}$
& \sotacell{98.44_{\pm 0.71}} \\

& F1 (w/o text)
& $86.97_{\pm 0.68}$
& $90.99_{\pm 0.58}$
& $\underline{95.61_{\pm 0.04}}$
& $91.54_{\pm 0.16}$
& --
& $87.29_{\pm 0.10}$
& $91.33_{\pm 0.31}$
& \sotacell{95.62_{\pm 0.61}} \\

& HM
& $65.32_{\pm 0.11}$
& $61.71_{\pm 1.06}$
& \sotacell{84.88_{\pm 0.43}}
& $78.20_{\pm 0.11}$
& $60.47_{\pm 0.47}$
& $65.83_{\pm 0.20}$
& $60.73_{\pm 0.15}$
& $\underline{80.87_{\pm 0.11}}$ \\

\addlinespace[2pt]
\multicolumn{2}{l}{
  \cellcolor{gainbg}\textit{Avg. Improv. of \our{}}
}
& \gaincell{+10.82\%}
& \gaincell{+14.91\%}
& \gaincell{-0.92\%}
& \gaincell{+12.83\%}
& \gaincell{+20.11\%}
& \gaincell{+9.53\%}
& \gaincell{+11.14\%}
& \cellcolor{gainbg}-- \\

\midrule

\multirow{4}{*}{Inter-table ($\uparrow$)}
& Walmart
& $78.12_{\pm 0.00}$
& $78.91_{\pm 1.17}$
& \sotacell{94.39_{\pm 0.46}}
& $79.62_{\pm 0.41}$
& $81.81_{\pm 1.37}$
& $84.50_{\pm 1.32}$
& $78.46_{\pm 3.23}$
& $\underline{90.65_{\pm 0.67}}$ \\

& Biodeg.
& $69.65_{\pm 0.00}$
& $72.09_{\pm 7.92}$
& $60.22_{\pm 1.82}$
& $64.68_{\pm 0.59}$
& --
& $63.68_{\pm 2.89}$
& $68.61_{\pm 0.87}$
& \sotacell{81.11_{\pm 12.18}} \\

& F1 (w/o text)
& $80.30_{\pm 0.54}$
& $82.47_{\pm 0.70}$
& $\underline{90.69_{\pm 0.26}}$
& $78.21_{\pm 1.12}$
& --
& $75.55_{\pm 0.90}$
& $83.01_{\pm 0.09}$
& \sotacell{92.04_{\pm 0.10}} \\

& HM
& $66.98_{\pm 2.45}$
& $66.13_{\pm 4.83}$
& $\underline{79.49_{\pm 6.18}}$
& $79.33_{\pm 0.10}$
& $73.47_{\pm 1.20}$
& $71.89_{\pm 2.62}$
& $70.90_{\pm 0.21}$
& \sotacell{83.15_{\pm 0.52}} \\

\addlinespace[2pt]
\multicolumn{2}{l}{
  \cellcolor{gainbg}\textit{Avg. Improv. of \our{}}
}
& \gaincell{+17.81\%}
& \gaincell{+16.18\%}
& \gaincell{+9.21\%}
& \gaincell{+15.44\%}
& \gaincell{+11.99\%}
& \gaincell{+18.03\%}
& \gaincell{+15.48\%}
& \cellcolor{gainbg}-- \\

\midrule

\multirow{4}{*}{CVR ($\downarrow$)}
& Walmart
& $1.085_{\pm 0.000}$
& $1.527_{\pm 0.083}$
& $0.649_{\pm 0.053}$
& $0.842_{\pm 0.080}$
& \sotacell{0.000_{\pm 0.000}}
& $0.685_{\pm 0.017}$
& $\underline{0.580_{\pm 0.050}}$
& \sotacell{0.000_{\pm 0.000}} \\

& Biodeg.
& $35.063_{\pm 0.008}$
& $32.038_{\pm 0.776}$
& $\underline{0.253_{\pm 0.065}}$
& $8.713_{\pm 0.186}$
& --
& $4.435_{\pm 0.293}$
& $6.571_{\pm 0.155}$
& \sotacell{0.006_{\pm 0.011}} \\

& F1 (w/o text)
& $44.416_{\pm 1.262}$
& $34.642_{\pm 3.265}$
& $\underline{31.346_{\pm 0.069}}$
& $31.442_{\pm 0.114}$
& --
& $33.744_{\pm 0.618}$
& $31.608_{\pm 0.227}$
& \sotacell{18.019_{\pm 0.606}} \\

& HM
& $23.240_{\pm 0.000}$
& $26.053_{\pm 2.331}$
& $11.735_{\pm 9.240}$
& $11.313_{\pm 2.713}$
& $34.573_{\pm 0.043}$
& $2.212_{\pm 0.031}$
& $\underline{1.696_{\pm 0.057}}$
& \sotacell{0.615_{\pm 0.168}} \\

\addlinespace[2pt]
\multicolumn{2}{l}{
  \cellcolor{gainbg}\textit{Avg. Improv. of \our{}}
}
& \gaincell{+89.19\%}
& \gaincell{+86.40\%}
& \gaincell{+83.72\%}
& \gaincell{+84.30\%}
& \gaincell{+49.11\%}
& \gaincell{+79.66\%}
& \gaincell{+76.65\%}
& \cellcolor{gainbg}-- \\

\midrule

TSTR (MAE) ($\downarrow$)
& Walmart
& $12805.23_{\pm 0.00}$
& $11601.78_{\pm 90.01}$
& $\underline{11345.76_{\pm 15.73}}$
& $11555.76_{\pm 16.86}$
& $11553.01_{\pm 6.93}$
& $11476.54_{\pm 104.06}$
& $11615.60_{\pm 29.02}$
& \sotacell{11318.90_{\pm 5.83}} \\

TSTR (MAE) ($\downarrow$)
& Biodeg.
& $1.1471_{\pm 0.0473}$
& $1.2487_{\pm 0.2471}$
& $1.1917_{\pm 0.0320}$
& $1.1796_{\pm 0.1129}$
& --
& $1.2518_{\pm 0.2388}$
& $\underline{1.1401_{\pm 0.0452}}$
& \sotacell{1.1250_{\pm 0.0574}} \\

TSTR (AUC) ($\uparrow$)
& F1 (w/o text)
& $0.5902_{\pm 0.1686}$
& $0.3560_{\pm 0.1142}$
& $\underline{0.7838_{\pm 0.0048}}$
& $0.3260_{\pm 0.0203}$
& --
& $0.7021_{\pm 0.0483}$
& $0.5130_{\pm 0.2037}$
& \sotacell{0.7935_{\pm 0.0052}} \\

TSTR (AUC) ($\uparrow$)
& HM
& $0.6266_{\pm 0.0366}$
& $0.6270_{\pm 0.0103}$
& $0.6098_{\pm 0.0282}$
& $0.6327_{\pm 0.0105}$
& $0.6158_{\pm 0.0133}$
& $\underline{0.6343_{\pm 0.0129}}$
& $0.6326_{\pm 0.0093}$
& \sotacell{0.6362_{\pm 0.0118}} \\

\addlinespace[2pt]
\multicolumn{2}{l}{
  \cellcolor{gainbg}\textit{Avg. Improv. of \our{}}
}
& \gaincell{+12.38\%}
& \gaincell{+34.18\%}
& \gaincell{+2.85\%}
& \gaincell{+37.66\%}
& \gaincell{+2.67\%}
& \gaincell{+6.21\%}
& \gaincell{+14.78\%}
& \cellcolor{gainbg}-- \\

\bottomrule
\end{tabular}%
}
\end{table*}

\begin{table}[t]
\centering
\caption{Synthesis quality on text-rich relational databases.
 Notation, highlighting, and average-improvement calculations follow Table~\ref{tab:quality_notext}.}
\label{tab:quality_text}
\small
\setlength{\tabcolsep}{3.8pt}
\renewcommand{\arraystretch}{0.95}
\resizebox{\columnwidth}{!}{%
\begin{tabular}{
@{}llcc
@{\hspace{4pt}\vrule width 0.45pt\hspace{4pt}}
c@{}
}
\toprule
Metrics & Datasets
& GReaT-Cond
& GReaT-Joint
& \our{} \\
\midrule

\multirow{3}{*}{C2ST-Agg ($\downarrow$)}
& Chicago
& $\underline{96.00_{\pm 0.28}}$
& $98.32_{\pm 0.42}$
& \sotacell{83.82_{\pm 0.35}} \\

& F1 (w/ text)
& $\underline{93.45_{\pm 0.13}}$
& $95.15_{\pm 0.56}$
& \sotacell{75.37_{\pm 0.93}} \\

& ML-small
& $\underline{92.04_{\pm 0.17}}$
& $98.44_{\pm 0.06}$
& \sotacell{78.93_{\pm 0.33}} \\

\addlinespace[2pt]
\multicolumn{2}{l}{
  \cellcolor{gainbg}\textit{Avg. Improv. of \our{}}
}
& \gaincell{+15.43\%}
& \gaincell{+18.45\%}
& \cellcolor{gainbg}-- \\

\midrule

\multirow{3}{*}{Intra-table ($\uparrow$)}
& Chicago
& $67.74_{\pm 1.01}$
& $\underline{72.60_{\pm 0.63}}$
& \sotacell{89.06_{\pm 0.18}} \\

& F1 (w/ text)
& $82.84_{\pm 0.52}$
& $\underline{86.83_{\pm 0.38}}$
& \sotacell{91.72_{\pm 0.44}} \\

& ML-small
& $\underline{97.49_{\pm 0.11}}$
& $96.51_{\pm 0.06}$
& \sotacell{98.96_{\pm 0.01}} \\

\addlinespace[2pt]
\multicolumn{2}{l}{
  \cellcolor{gainbg}\textit{Avg. Improv. of \our{}}
}
& \gaincell{+14.57\%}
& \gaincell{+10.28\%}
& \cellcolor{gainbg}-- \\

\midrule

\multirow{3}{*}{Inter-table ($\uparrow$)}
& Chicago
& $67.69_{\pm 1.93}$
& $\underline{71.67_{\pm 0.91}}$
& \sotacell{81.41_{\pm 0.31}} \\

& F1 (w/ text)
& $85.15_{\pm 0.30}$
& $\underline{90.02_{\pm 0.10}}$
& \sotacell{92.94_{\pm 0.38}} \\

& ML-small
& $\underline{95.08_{\pm 0.29}}$
& $94.51_{\pm 0.17}$
& \sotacell{98.24_{\pm 0.14}} \\

\addlinespace[2pt]
\multicolumn{2}{l}{
  \cellcolor{gainbg}\textit{Avg. Improv. of \our{}}
}
& \gaincell{+10.91\%}
& \gaincell{+6.93\%}
& \cellcolor{gainbg}-- \\

\midrule

\multirow{3}{*}{CVR ($\downarrow$)}
& Chicago
& $\underline{98.946_{\pm 0.381}}$
& $99.411_{\pm 0.005}$
& \sotacell{21.017_{\pm 0.306}} \\

& F1 (w/ text)
& $\underline{32.529_{\pm 0.502}}$
& $36.938_{\pm 0.318}$
& \sotacell{12.397_{\pm 1.020}} \\

& ML-small
& $\underline{0.829_{\pm 0.036}}$
& $45.951_{\pm 37.317}$
& \sotacell{0.809_{\pm 0.139}} \\

\addlinespace[2pt]
\multicolumn{2}{l}{
  \cellcolor{gainbg}\textit{Avg. Improv. of \our{}}
}
& \gaincell{+47.69\%}
& \gaincell{+81.18\%}
& \cellcolor{gainbg}-- \\

\midrule

TSTR (AUC) ($\uparrow$)
& Chicago
& $0.4946_{\pm 0.0203}$
& $\underline{0.6351_{\pm 0.1267}}$
& \sotacell{0.6815_{\pm 0.0115}} \\

TSTR (AUC) ($\uparrow$)
& F1 (w/ text)
& $\underline{0.7899_{\pm 0.0358}}$
& $0.5613_{\pm 0.0669}$
& \sotacell{0.8159_{\pm 0.0424}} \\

TSTR (MAE) ($\downarrow$)
& ML-small
& $0.9148_{\pm 0.0091}$
& \sotacell{0.8917_{\pm 0.0008}}
& $\underline{0.9122_{\pm 0.0089}}$ \\

\addlinespace[2pt]
\multicolumn{2}{l}{
  \cellcolor{gainbg}\textit{Avg. Improv. of \our{}}
}
& \gaincell{+13.79\%}
& \gaincell{+16.79\%}
& \cellcolor{gainbg}-- \\

\bottomrule
\end{tabular}%
}
\end{table}

\subsection{Synthesis Quality}
\label{sec:exp_quality}

\noindent\textbf{Exp-1: Quality on text-free relational databases.}
Table~\ref{tab:quality_notext} compares \our{} with all baselines on four text-free relational databases.
The results show that \our{} achieves strong performance across all evaluated metrics. Specifically, \our{} achieves the lowest C2ST-Agg on three databases and only trails GRDM on Walmart. Its average reduction in C2ST-Agg is at least 7.36\% over the baselines, indicating that the synthetic records are more realistic. Regarding pairwise relationships, \our{} ranks first on two databases and second on the other two in intra-table similarity. Its inter-table similarity is even stronger, ranking first on three databases and second on Walmart, with an average improvement of at least 9.21\% over the baselines. These results demonstrate that \our{} effectively preserves relationships between columns both within and across tables.
In terms of semantic consistency, \our{} achieves the lowest CVR on all four databases. The average relative reduction in CVR is at least 49.11\% over the baselines on their applicable databases, reflecting better adherence to the verified dependencies. Finally, \our{} achieves the best TSTR performance on all four downstream tasks, suggesting that its synthetic data supports downstream training.

\noindent\textbf{Exp-2: Quality on text-rich relational databases.}
Table~\ref{tab:quality_text} compares \our{} with GReaT-Cond and GReaT-Joint on three text-rich relational databases.
The results show that \our{} achieves the best performance across all databases and metrics, except for TSTR on ML-small, where its MAE is 2.30\% higher than that of GReaT-Joint. On average, \our{} reduces C2ST-Agg by at least 15.43\% and improves intra-table and inter-table similarities by at least 10.28\% and 6.93\%, respectively, over the baselines. It also achieves an average  reduction of at least 47.69\% in CVR and an average  improvement of at least 13.79\% in downstream utility. These results show gains in fidelity and semantic consistency alongside improved average downstream utility. They support a unified framework for jointly synthesizing structured values and text across related tables.

\subsection{Quality-Efficiency Trade-off}
\label{sec:exp_efficiency}

\begin{figure}[t]
\centering
\includegraphics[width=0.98\columnwidth]{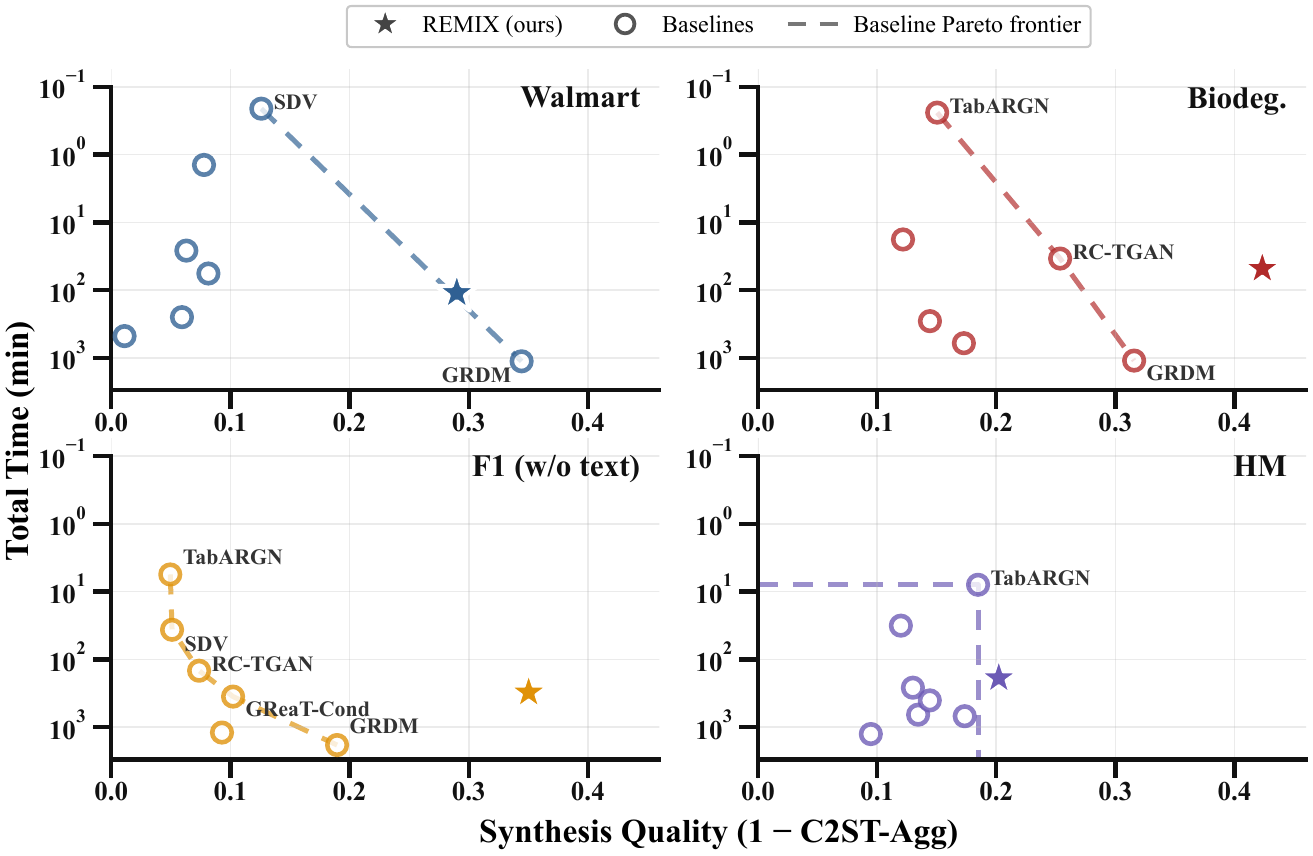}
\caption{Quality-efficiency trade-off on text-free relational databases. Synthesis quality is measured by $1-\mathrm{C2ST\text{-}Agg}$, and total time includes training and generation.}
\label{fig:efficiency}
\end{figure}

\noindent\textbf{Exp-3: Quality-efficiency trade-off.}
In this experiment, we evaluate the quality-efficiency trade-off on four text-free relational databases. We use $1-\mathrm{C2ST\text{-}Agg}$ to measure synthesis quality and record the total training and generation time. We plot the Pareto frontier of the baselines on each database to examine whether \our{} improves the existing balance between quality and efficiency. The results are presented in Figure~\ref{fig:efficiency}.
As shown in the figure, \our{} extends the baseline Pareto frontier on all four databases. Specifically, fast baselines such as SDV, RC-TGAN, and TabARGN offer lower time costs on some databases, but consistently yield lower synthesis quality than \our{}. Compared with GRDM, \our{} is 22.8$\times$, 6.0$\times$, and 3.7$\times$ faster on Biodeg., F1 (w/o text), and HM, respectively, while achieving higher synthesis quality. On Walmart, although GRDM achieves higher quality, \our{} is 10.2$\times$ faster, providing strong synthesis quality at a substantially lower time cost. Collectively, these results show that \our{} achieves a superior balance between synthesis quality and efficiency across the evaluated databases.

\subsection{Ablation Study}
\label{sec:exp_ablation}

\begin{table*}[t]
\centering
\caption{
Ablation study of \our{} on synthesis fidelity.
C2ST-Agg ($\downarrow$, \%) is reported as mean $\pm$
standard deviation over three runs.
Avg. Improv. denotes the mean absolute reduction in C2ST-Agg
achieved by the full model over each ablated variant.
}
\label{tab:ablation_study}
\scriptsize
\setlength{\tabcolsep}{4.2pt}
\renewcommand{\arraystretch}{1.05}
\resizebox{\textwidth}{!}{%
\begin{tabular}{@{}lcccccccc@{}}
\toprule
Variant
& Walmart
& Biodeg.
& F1 (w/o text)
& HM
& Chicago
& ML-small
& F1 (w/ text)
& Avg. Improv. \\
\midrule
\our{}
& $71.85 \pm 3.57$
& $57.67 \pm 0.45$
& $64.97 \pm 0.50$
& $79.79 \pm 0.29$
& $83.82 \pm 0.35$
& $78.93 \pm 0.33$
& $75.37 \pm 0.93$
& -- \\

w/o Encoder
& $89.63 \pm 3.57$
& $57.35 \pm 0.28$
& $74.12 \pm 0.72$
& $89.74 \pm 0.12$
& $78.21 \pm 0.79$
& $78.12 \pm 0.64$
& $87.07 \pm 0.61$
& $+5.98\%$ \\

w/o GNN
& $69.26 \pm 5.01$
& $56.68 \pm 0.60$
& $74.52 \pm 0.23$
& $83.04 \pm 0.58$
& $89.94 \pm 0.98$
& $85.38 \pm 0.05$
& $83.24 \pm 0.91$
& $+4.24\%$ \\

w/o Residual
& $67.41 \pm 4.21$
& $66.97 \pm 0.78$
& $70.59 \pm 1.36$
& $85.89 \pm 0.29$
& $81.24 \pm 0.77$
& $86.46 \pm 0.47$
& $82.94 \pm 0.26$
& $+4.16\%$ \\
\bottomrule
\end{tabular}%
}
\end{table*}

\noindent\textbf{Exp-4: Ablation study.}
In this experiment, we conduct an ablation study to evaluate the effectiveness of three key modules in \our{}, including the type-aware row encoder, the heterogeneous GNN, and the stochastic row residual. To evaluate the impact of the type-aware row encoder, we replace it with direct row serialization followed by SLM embedding while keeping the other components unchanged. To assess the effectiveness of the GNN-enhanced representation, we remove the heterogeneous GNN and directly project the dropout-masked row representations into soft prompt tokens. Lastly, we remove the stochastic row residual and project only the GNN-enhanced representation into soft prompt tokens. The results are summarized in Table~\ref{tab:ablation_study}.
As shown, type-aware encoding reduces C2ST-Agg by 5.98\%, highlighting its value in constructing informative representations for records with heterogeneous attributes. The GNN-enhanced representation contributes an average reduction of 4.24\%, demonstrating the benefit of incorporating relational context into generation. Retaining the stochastic row residual lowers C2ST-Agg by 4.16\% on average, indicating that row-level information complements the GNN-enhanced representation. These results collectively demonstrate the effectiveness of the three modules in improving synthesis fidelity.

\subsection{Hyperparameter Analysis}
\label{sec:exp_hyperparameters}

\begin{figure*}[t]
    \centering
    \setlength{\tabcolsep}{0pt}
    \renewcommand{\arraystretch}{0.2}

    \begin{tabular}{@{}c@{\hspace{0.02\textwidth}}c@{}}
        \includegraphics[width=0.46\textwidth]{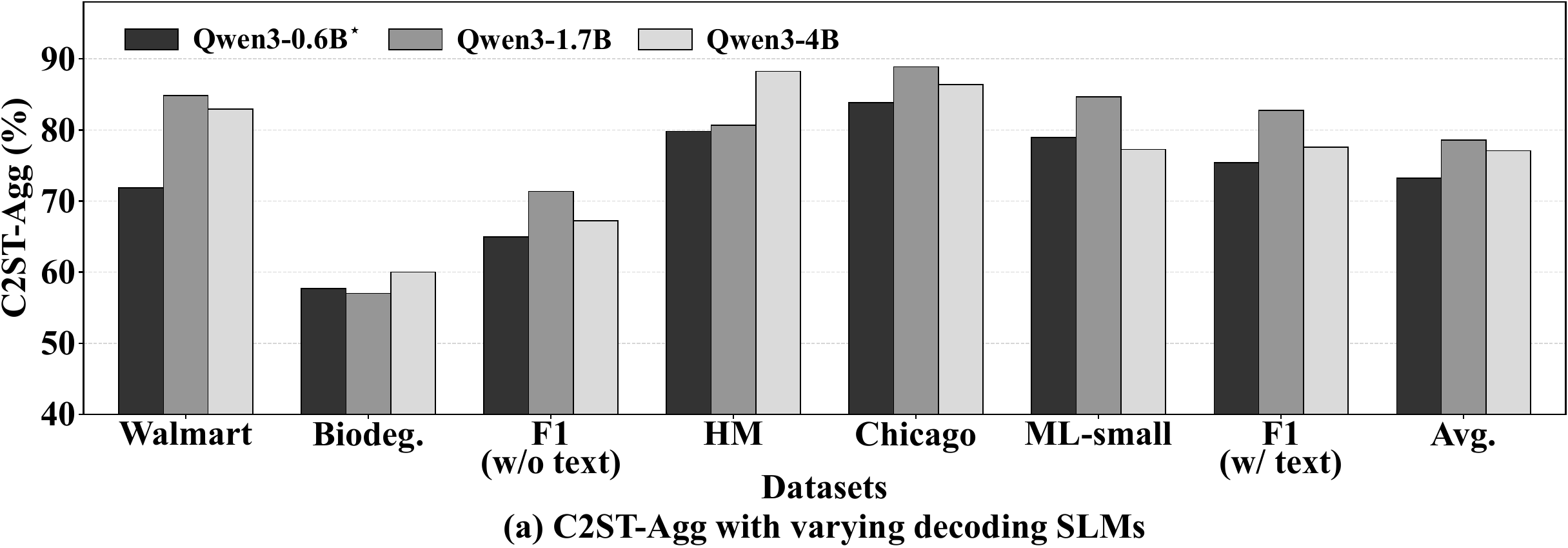}
        &
        \includegraphics[width=0.46\textwidth]{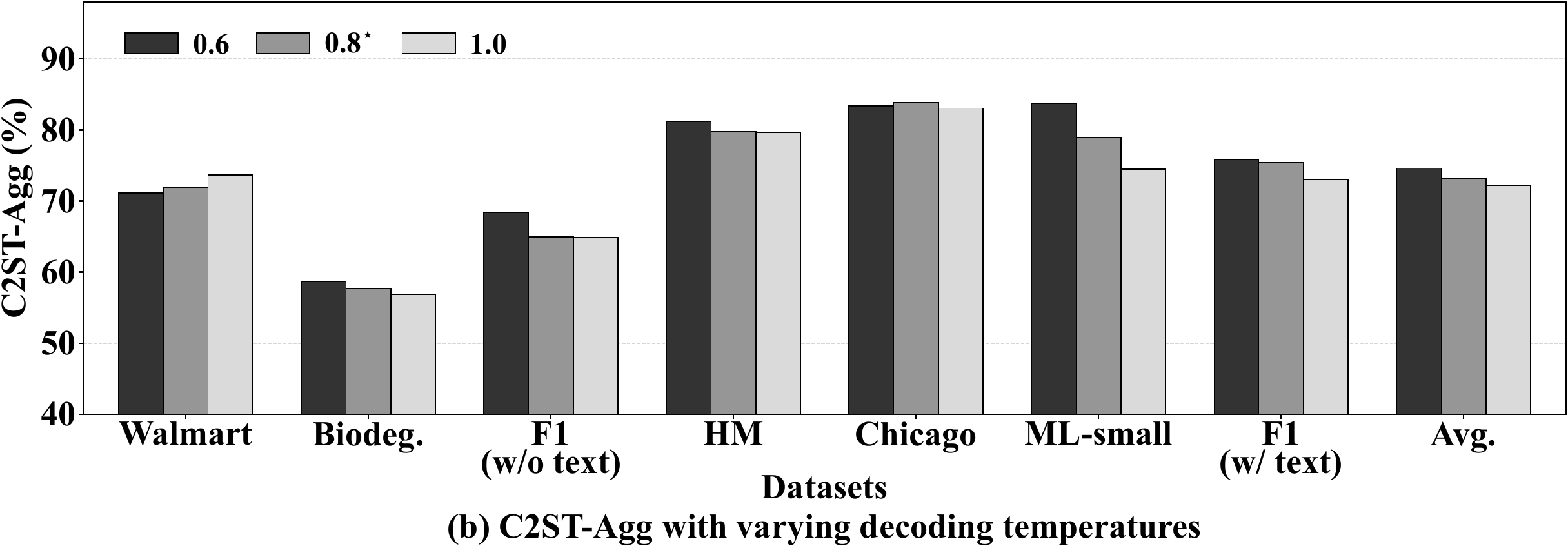}
        \\[2pt]
        \includegraphics[width=0.46\textwidth]{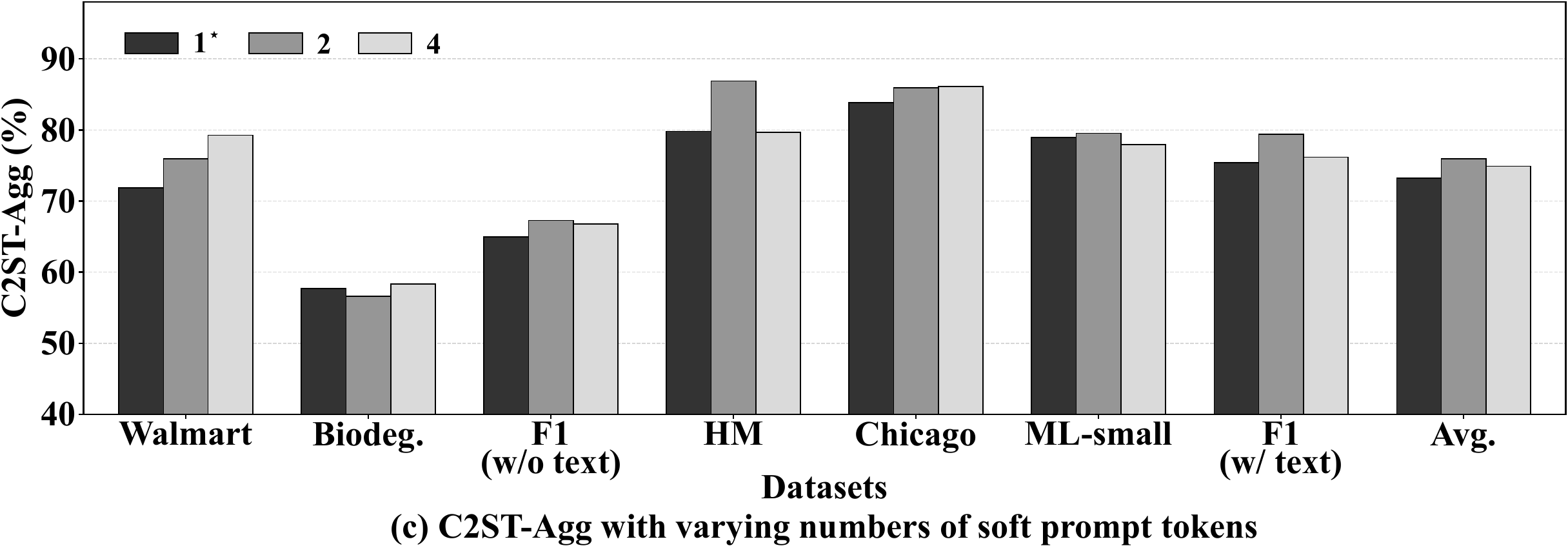}
        &
        \includegraphics[width=0.46\textwidth]{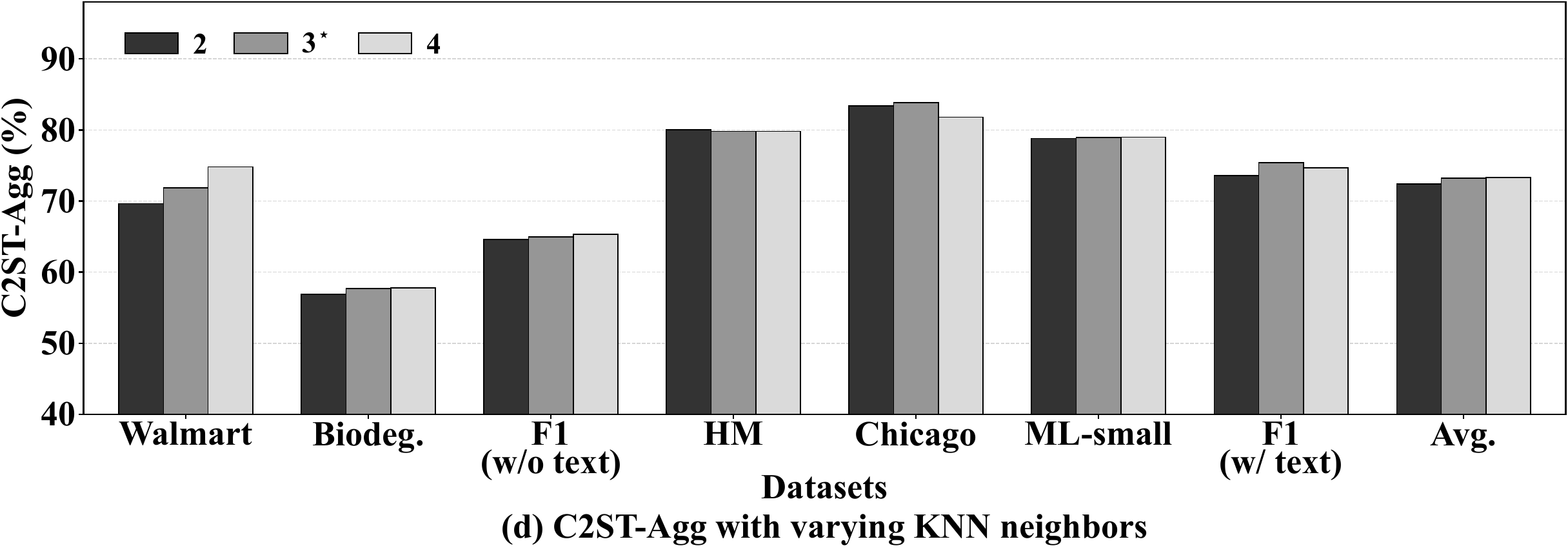}
    \end{tabular}

    \caption{
Hyperparameter sensitivity of \our{}.
Panels (a)--(d) vary decoder size, decoding temperature,
the number of soft prompt tokens, and the number of nearest
neighbors ($k$), respectively.
Results show C2ST-Agg ($\downarrow$, \%) for each database
and the mean across databases.
Stars mark the default settings.
}
    \label{fig:hyperparameters}
\end{figure*}

\noindent\textbf{Exp-5: Varying decoding SLMs.}
We evaluate \our{} with three decoder sizes from the Qwen3 family: Qwen3-0.6B, Qwen3-1.7B, and Qwen3-4B. The results are presented in Figure~\ref{fig:hyperparameters}(a). As shown, increasing the decoder size does not consistently improve fidelity. Qwen3-0.6B achieves the lowest C2ST-Agg on five of the seven databases and the best average of 73.20\%. Although Qwen3-1.7B improves fidelity on Biodeg.\ and Qwen3-4B performs better on ML-small, these gains do not offset their degradation on the other databases. These results suggest that larger decoder capacity alone does not guarantee better synthesis fidelity. Therefore, we use Qwen3-0.6B as the default decoder.

\noindent\textbf{Exp-6: Varying decoding temperatures.}
We evaluate \our{} under decoding temperatures of 0.6, 0.8, and 1.0. Figure~\ref{fig:hyperparameters}(b) summarizes the results. As illustrated, increasing the temperature generally improves fidelity, although the changes are moderate: the average C2ST-Agg varies by only 2.37\% across the tested settings. Overall, \our{} maintains strong synthesis fidelity while different databases exhibit different temperature preferences. We therefore adopt 0.8 as the default temperature, choosing an intermediate setting that provides strong overall performance.

\noindent\textbf{Exp-7: Varying the number of soft prompt tokens.}
We evaluate \our{} with 1, 2, and 4 soft prompt tokens. As shown in Figure~\ref{fig:hyperparameters}(c), using more tokens does not consistently improve fidelity. One token achieves the lowest average C2ST-Agg of 73.20\%, compared with 75.94\% and 74.89\% for two and four tokens, respectively. Although additional tokens bring gains on a few databases, they provide no overall improvement. This suggests that one soft prompt token is sufficient to convey conditioning information to the decoder. Therefore, we use one token as the default setting.

\noindent\textbf{Exp-8: Varying the number of nearest neighbors ($k$).}
We vary the number of same-table nearest neighbors used to aggregate soft prompts during synthesis, considering $k=2$, $3$, and $4$. The results are reported in Figure~\ref{fig:hyperparameters}(d). Across these settings, \our{} maintains strong synthesis fidelity, suggesting that its performance is relatively insensitive to the choice of $k$. Different databases exhibit different preferences for the number of neighbors. We therefore adopt $k=3$ as the default setting, as it provides consistently strong fidelity across databases.

\subsection{Semantic Dependency Engine Evaluation}
\label{sec:exp_constraint_engine}

\begin{figure}[t]
\centering
\includegraphics[width=0.95\columnwidth]{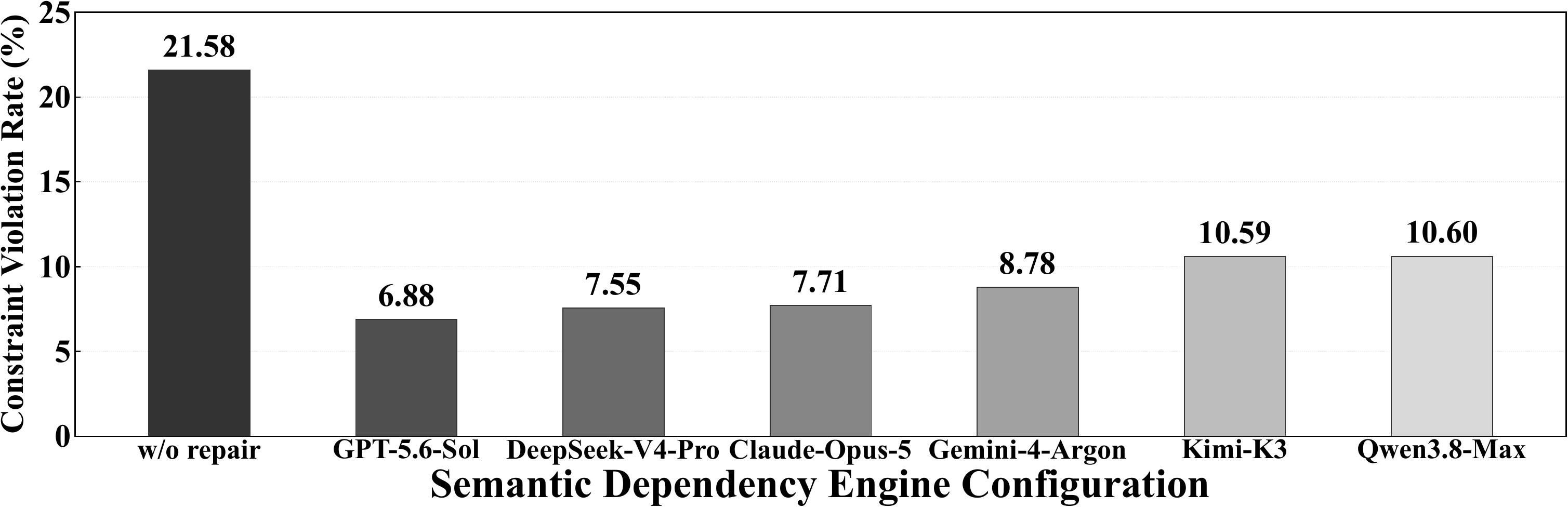}
\caption{
Mean CVR ($\downarrow$, \%) under different semantic dependency engine
configurations over three generation runs, each averaged equally across
seven databases.
}
\label{fig:constraint_engine}
\end{figure}

\noindent\textbf{Exp-9: Dependency engine analysis.}
We evaluate the effectiveness of the semantic dependency engine and its robustness across different LLMs. The results are summarized in Figure~\ref{fig:constraint_engine}. Without semantic dependency enforcement, the average CVR reaches 21.58\%. Our evaluation covers three frontier proprietary models, namely GPT-5.6-Sol, Claude-Opus-5, and Gemini-4-Argon, and three leading open-source models, namely DeepSeek-V4-Pro, Kimi-K3, and Qwen3.8-Max. With the semantic dependency engine, the three proprietary models lower the average CVR to 6.88\%, 7.71\%, and 8.78\%, respectively, while the three open-source models achieve 7.55\%, 10.59\%, and 10.60\%, respectively. All six configurations reduce CVR by more than 50\%, demonstrating the engine's effectiveness and robustness across the evaluated LLMs, although the extent of improvement varies with the underlying model. Considering reproducibility requirements and performance, we select the open-source DeepSeek-V4-Pro as the default LLM.

%% file: relatedwork.tex
\section{Related Work}
\label{sec:related_work}

Existing relational database synthesis methods can be broadly categorized into four groups: statistical methods, conventional learning-based methods, diffusion-based methods, and LLM-based methods.

\noindent\textbf{Statistical methods.}
Statistical methods explicitly model attribute distributions and inter-table relationships. The original SDV~\cite{sdv} combines Gaussian copulas with hierarchical modeling, propagating statistics of child groups into parent representations before sampling related tables. PrivLava~\cite{privlava} introduces latent variables to capture correlations among tuples sharing foreign-key references and synthesizes relational data under differential privacy.

\noindent\textbf{Conventional learning-based methods.}
Conventional learning-based methods learn attribute distributions and inter-table dependencies using GANs~\cite{gan}, VAEs~\cite{vae}, or autoregressive models~\cite{ar}. RC-TGAN~\cite{rctgan} adapts CTGAN~\cite{ctgan} to relational database synthesis by conditioning child-table generators on parent rows. GraphVAE~\cite{graphvae} combines graph neural networks with variational autoencoders to synthesize attributes while retaining the input relational graph structure. NeuralM2M~\cite{m2m} first generates a graph representing relationships between records across tables, then synthesizes attributes using neural autoregressive models conditioned on node embeddings and neighboring records. TabularARGN~\cite{argn} follows a parent-to-child generation order, synthesizing child sequences with an autoregressive network conditioned on parent attributes and LSTM-encoded history.

\noindent\textbf{Diffusion-based methods.}
Diffusion-based methods generate row attributes through iterative denoising conditioned on relational information. ClavaDDPM~\cite{ddpm} jointly generates parent records and cluster-derived latent variables with diffusion, then uses these variables to guide child-row diffusion. RGCLD~\cite{rgcld} generates tables sequentially with latent diffusion models conditioned on GNN embeddings. In contrast, GRDM~\cite{grdm} and RelDiff~\cite{reldiff} first generate a graph representing the database structure, then jointly denoise row attributes across tables using heterogeneous GNNs that aggregate information from connected records at each diffusion step.
These methods primarily model numerical and categorical attributes, whereas \our{} combines graph representations with language modeling to jointly synthesize structured values and text.

\noindent\textbf{LLM-based methods.}
LLM-based methods serialize records into token sequences and synthesize data using language-model architectures. REaLTabFormer~\cite{realtabformer} generates parent records with GPT-2~\cite{gpt2}, then uses a sequence-to-sequence model to generate child records conditioned on their parents. GReaTER~\cite{greater} restructures related tables by joining correlated features and removing duplicate records. It also enhances the semantics of categorical labels before synthesizing relational databases using GPT-2-based models.
These methods obtain relational context from serialized parent records or restructured tables, whereas \our{} aggregates information from connected records into soft prompts.
It further applies tested computation and repair programs derived from validated semantic dependencies to reduce inconsistencies.

%% file: conclusion.tex
\section{CONCLUSION}
\label{sec:conclusion}

In this paper, we study mixed-type relational database synthesis and propose \our{}, a unified framework based on collaboration between an SLM and an LLM. \emph{Graph-to-Prompt Tuning} jointly learns graph-conditioned soft prompts and tunes the SLM through attribute reconstruction. \emph{Collaborative SLM--LLM Decoding} combines SLM generation with tested computation and repair programs derived from LLM-discovered, independently validated semantic dependencies. Experiments on seven real-world relational databases demonstrate the superiority of \our{}. Future work will explore scaling \our{} to larger relational databases and applying it to more downstream machine learning tasks.